\documentclass[number]{cas-sc}

\usepackage[numbers]{natbib}
\usepackage[utf8]{inputenc}
\usepackage{amsmath}
\usepackage[ruled,vlined]{algorithm2e} 
\usepackage{array}          
\usepackage{tabularx}       
\usepackage{booktabs}        
\usepackage{graphicx}       
\usepackage{float}
\usepackage{graphicx}
\usepackage{subcaption}
\usepackage{booktabs}
\usepackage{siunitx}

\def\tsc#1{\csdef{#1}{\textsc{\lowercase{#1}}\xspace}}
\tsc{WGM}
\tsc{QE}
\tsc{EP}
\tsc{PMS}
\tsc{BEC}
\tsc{DE}

\begin{document}
\let\WriteBookmarks\relax
\def\floatpagepagefraction{1}
\def\textpagefraction{.001}

\shorttitle{ExPUFFIN}

\shortauthors{REBELLO, C. M. and DI CAPRIO, U, et~al.}

\title [mode = title]{ExPUFFIN: Thermodynamic Consistent Viscosity Prediction in an Extended Path-Unifying Feed-Forward Interfaced Network}                      



%
\author[1]{Carine Menezes Rebello}[orcid=0000-0002-0796-8116]
\ead{carine.m.rebello@ntnu.no}

\fnmark[1]
\cormark[1]
\fntext[eq]{These authors share first authorship.}

\author[2]{Ulderico {Di Caprio}}[orcid=0000-0001-5194-8721]
\fnmark[1]

\ead{ulderico.dicaprio@kuleuven.be}

\author[1]{Jenny Steen-Hansen}

\author[1]{Bruno Rodrigues}[orcid=0009-0006-0513-3699]

\author[1]{Erbet Almeida Costa}[orcid=0000-0003-1397-9628]

\author[3]{Anderson Rapello dos Santos}

\author[2]{Flora Esposito}[orcid=0009-0008-9396-1922]
                       
\author[2]{Mumin Enis Leblebici}[orcid=0000-0003-4599-9412]

\author[1]{Idelfonso B. R. Nogueira}[orcid=0000-0002-0963-6449]

\affiliation[1]{organization={Department of Chemical Engineering},
    addressline={Norwegian University of Science and Technology}, 
    city={Trondheim},
    postcode={793101}, 
    country={Norway}}

\affiliation[2]{organization={Process Engineering for Sustainable Systems, Department of Chemical Engineering},
    addressline={KU Leuven, Agoralaan Building B}, 
    city={Diepenbeek},
    postcode={3590}, 
    country={Belgium}}

\affiliation[2]{organization={Petróleo Brasileiro S.A. (PETROBRAS), Wells/Well Engineering},
    city={Rio de Janeiro},
    country={Brazil}}

\cormark[1]
\cortext[cor1]{Corresponding author}



\begin{abstract}
Accurate prediction of liquid viscosity is essential for process design and simulation, yet remains challenging for novel molecules. Conventional group-contribution models struggle with isomer discrimination, large molecules, and parameter availability, while purely data-driven graph neural networks (GNNs) demand large datasets and offer limited interpretability. Even when feasible to be applied, purely data-driven models lack thermodynamic consistency in their predictions and are not a reliable solution. This work introduces ExPUFFIN, an extended version of the Path-unifying Feed-Forward Interfaced Network, consisting of a hybrid GNN-based framework that directly predicts temperature-dependent viscosities of pure hydrocarbons from molecular graphs, while enforcing mechanistic inductive biases in the output layer to ensure thermodynamic consistency. Molecular information is given as graph structures, encoded as a graph convolutional network, and mapped to an inductive bias neuron based on two thermophysical correlations: a three-parameter Andrade-type equation and a four-parameter empirical viscosity–temperature relation. The accuracy of these models is compared with a solely data-driven prediction. The Andrade-based ExPUFFIN variant reduces RMSE compared to the purely data-driven baseline of 37\% and yields smooth, physically consistent interpolation and extrapolation of viscosity–temperature curves, properties that are not observed in purely data-driven models. The empirical ExPUFFIN model provides comparable accuracy while retaining robust trends. Overall, embedding physics-based structure in GNN outputs improves accuracy, robustness, and transferability, enabling reliable viscosity predictions for complex hydrocarbon molecules. The approach is readily extendable to other properties and significantly broader chemical domains.
\end{abstract}


\begin{highlights}
\item ExPUFFIN combines GNNs with physical-inductive models to improve viscosity predictions.
\item The hybrid approach generates smoother and more physically coherent viscosity-temperature curves.  
\item The ExPUFFIN model improves interpolation and extrapolation when compared to the baseline model.
\end{highlights}


\maketitle

\section{Introduction}
Viscosity is a property that plays an important role in various engineering applications, including process design, simulation, optimisation, and control of systems involving fluid flow and transport phenomena. Accurate viscosity data are essential in the formulation of fuels, oil transport, lubricants, pharmaceuticals, and specialty chemicals, impacting product quality, process efficiency, and safety. Despite its importance, the reliable prediction of viscosity for novel or hypothetical molecules remains challenging. 

Available methods for predicting the viscosity of novel molecules rely on Quantitative structure-property relationship (QSPR). These models correlate the predicted property with the molecular structure. Furthermore, they allow for including dependence of the property on the system-level variables (e.g., temperature and pressure). Representing molecular structure in these techniques is a common challenge. One of the most used approaches to accomplish this task is the group contribution method \cite{gani2019}. Here, the molecular is approximated to a mixture of known molecular groups, each defined by standard model parameters. The group contribution method has been widely used to predict viscosity, both using physical predictors \cite{cardonaetal2021} and machine learning (ML) approaches \citep{goussardetal2020, roostaetal2023}. However, despite the significant advantages of these techniques, they have many limitations, including the inability to distinguish between isomers, limited precision for large molecules, and the inaccessibility of certain parameters within the molecule \cite{gani2019}. 

In recent years, graph neural networks (GNNs) have emerged as a valid alternative to overcome group contribution limitations. GNNs are ML models able to observe the entire structure of the molecule and infer from it the characteristics needed to predict the targeted molecular property through a supervised training approach. GNNs are based on classic graph theory, where a graph is represented numerically by building the adjacency and feature matrices. For chemical molecules, atoms and bonds are represented as nodes and edges of a graph, each with its own properties (e.g., molecular weight and bond type), enabling direct learning from molecular topology without manually engineered descriptors \cite{Rittigetalbook2023}. In this sense, graph theory is a straightforward way to include domain-specific information, in this case chemistry, into data-driven models. 

In the domain of chemical property prediction, GNNs have been used to predict fuel ignition properties \citep{leenhoutsetal2025, schweidtmannetal2020}, solvation-free energies \cite{vermeireetal2021}, activity coefficients \cite{rittigetal2023}, vapour pressure \cite{vienasantanaetal2024}, and further pure compound properties \cite{aouichaouietal2023}. Showing high accuracy and dictating new standards within the state of the art of property predictions. GNNs have also been applied in the domain of viscosity predictions for both pure compounds \cite{chewetal2024} and binary mixtures \cite{bilodeauetal2023}, showing outstanding prediction capabilities. Despite the significant advantages and high accuracy of GNNs, they demand large datasets to be trained because the model needs to infer all the input-output correlations solely relying on data. Therefore, even though its inputs are intrinsically domain-aware, inherited by the employment of graph theory and atomic structure, its outputs are not physically bounded, inheriting the limitations of traditional data-driven models (e.g., poor extrapolability, lack of physical consistency, and limited interoperability for sparse data).

Hybrid modelling (HM) and physics-informed machine learning (PI-ML) are ML techniques that overcome the issues of pure GNN approaches. In fact, they combine mechanistic information about the system with data-driven knowledge, decreasing the amount of data needed for the training and increasing the system's interpretability and generalization capabilities \cite{schweidtmannetal2024, espositohmpi2025, vonstoschhmreview}. They do not need to infer all the information from the data, since it is already partially represented in the mechanistic model, which eases comprehension of the model's decision pattern. The most common application of hybrid modelling lies where a consolidated and systematic mathematical representation is available for the phenomena under study, with a recognizable level of uncertainty connected to part of that mathematical representation. For example, in mass, energy, and momentum balances where the sink/source term has a high level of uncertainty, and the remaining constitutive terms are connected with well-established conservation principles. \cite{santanaetal2023, nogueiraetal2022, dicaprioetal2023, vonstoschhmreview, jesperetal_2025, destroetal2020}

In the domain of predicting pure compound or mixture properties, the hybridization approach is not so straightforward, and the field is still under development, with increased interest in how to develop PI-ML approaches for such applications. For instance, Viena et al. enforced the Antoine equation in the last layer of a GNN predicting vapour pressure for pure compounds \cite{vienasantanaetal2024}, and Rittig et al. enforced the Gibbs-Duhem equation for predicting activity coefficients in binary mixtures \cite{rittigetal2023}. However, such an application lacks the systematic mathematical formalisation and clear distinction between well-known and uncertain phenomena, as aforementioned. Thus, despite the significant advantages brought by model hybridization, the literature still lacks a systematic methodology to enforce known correlations in GNN when predicting physical properties, as is the case for the viscosity of pure chemical species. Where, to our knowledge, the only available work using PI-ML is that of Chew et al., who focused on hybridizing the model in the input space by selecting molecular dynamics descriptors to be used as inputs to enhance model accuracy and explainability \cite{chewetal2024}. On the other hand, informing the output layer with known correlations between the viscosity and the process parameters (i.e., liquid temperature) would be beneficial for both increasing model accuracy and interpretability, while facilitating easy transferability to practitioners who are not experts in ML for novel molecules. Moreover, this would address the previously mentioned limitations of purely data-driven models.

This paper proposes a novel QSPR model leveraging GNNs to predict the viscosity of pure compounds and their dependence on temperature, solely relying on their graph representation and temperature. The methodology proposes a hybrid modelling approach that, in contrast to the usual hybridisation of a constitutive system mathematical representation, hybridizes the structure of a purely data-driven model, combining the learning capabilities of ML techniques and the robustness of mechanistic models. More specifically, two inductive biases are enforced to predict the viscosity, namely the Andrade model and an empirical correlation employed in the literature. These models are enforced in the last model layer to increase the physical adherence to the learned information. Furthermore, it leverages the domain information inherited from graph theory applied to chemical components, so both inputs and outputs are domain-informed. 

Accordingly, we propose a modified version of a GNN that systematically incorporates chemical and thermodynamic domain information in both the molecular input representation and the viscosity output layer for temperature-dependent viscosity prediction. By constraining the model in this way, we expect to improve the reliability of the data-driven predictions in interpolation and extrapolation across temperature ranges. Additionally, such a configuration allows mapping each molecule using a well-defined set of parameters, increasing the transferability of the resulting knowledge to non-experts in ML and enhancing the impact these tools can have on process development. This approach we call ExPUFFIN, as a natural extention of the concepts of Path-Unifying Feed-Forward Interfaced Network \cite{vienasantanaetal2024}.

\section{Methodology}
\begin{figure}[h!]
	\centering
		\includegraphics[scale=.12]{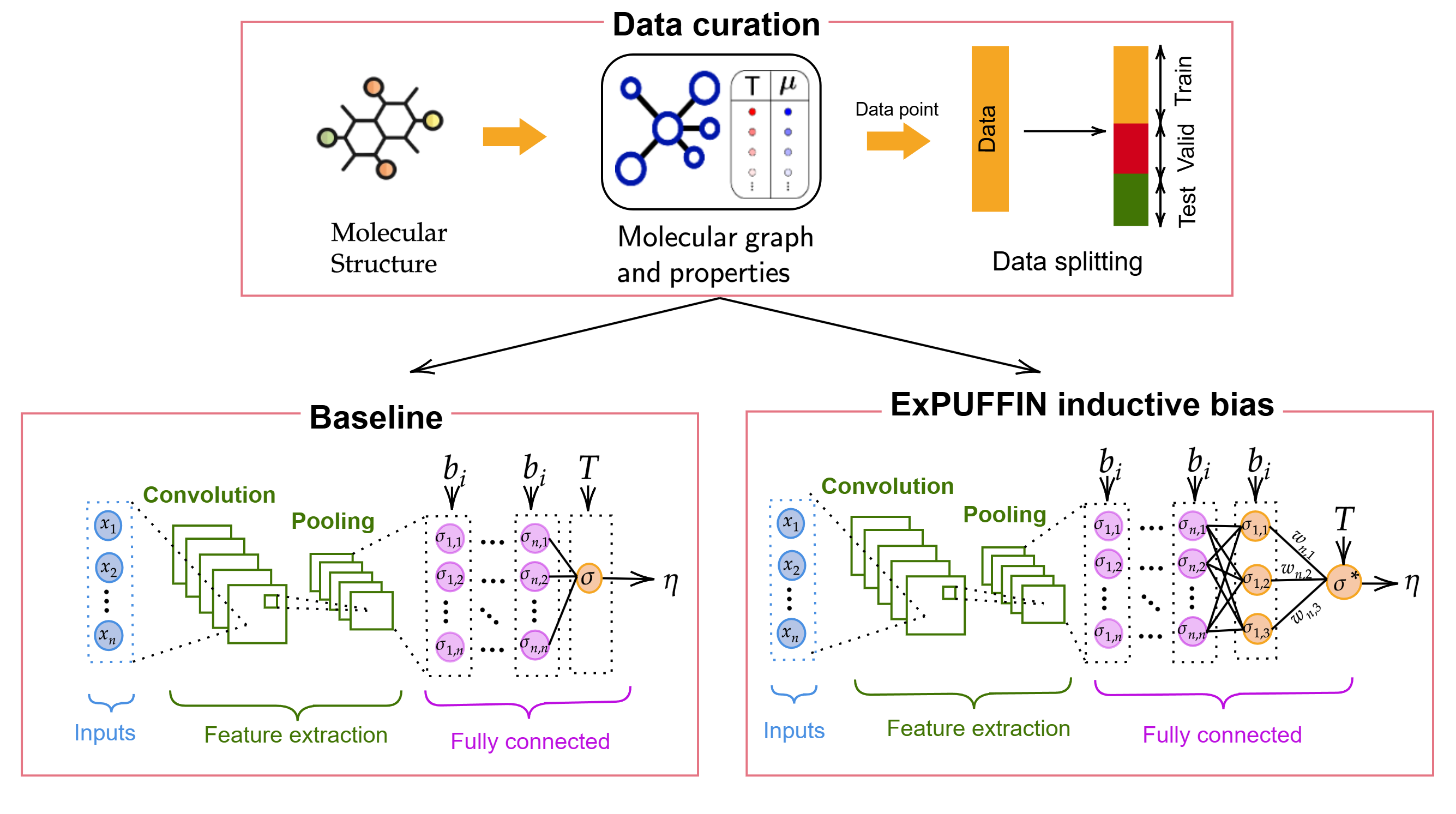}
	\caption{Methodology pipeline illustrating data curation, dataset splitting, and model architectures.}
	\label{fig:Methodology}
 \end{figure}

In this work, we have applied graph neural networks (GNNs) to predict the temperature dependence of viscosity, leveraging inductive bias techniques. The adopted methodology is described in Figure \ref{fig:Methodology}. The molecules are initially represented by their chemical structures, converted into molecular graphs, and associated with their respective experimental viscosities through a data curation process. After organizing and dividing the data into training, validation, and test sets, three types of models are employed: 1) a baseline model, which directly learns the relationship between molecular structure and viscosity, 2) a model with a physical inductive bias, which incorporates the functional form of the Andrade equations into the prediction process, and 3) a model with a physical inductive bias, which incorporates the functional form of empirical equations into the prediction process.
In Section \ref{data_curation}, we describe how the data have been obtained and curated toward the various dataset preparations. In Section \ref{model_tail}, we describe the employed GNN and the featurisation employed for the molecular graph generation. In Section \ref{model_head}, we describe the employed baseline model and inductive biases. In Section \ref{model_and_training_details}, we describe the details of the model structure and training.

\subsection{Data curation} \label{data_curation}

In this study, an approach based on structural molecular representations was adopted to model and predict the properties of chemical compounds. The accuracy of such predictive solutions depends directly on how molecules are numerically described, making the representation step a central component of the modeling process.

Molecular systems can be encoded using different structural formats that preserve connectivity patterns, geometric arrangements, or electronic structure. Common representations in data-driven modeling include molecular graphs, adjacency matrices, InChI strings, and the simplified molecular input line entry specification (SMILES). Among these, SMILES strings stand out for their compactness, expressiveness, and compatibility with graph-based learning.

\begin{figure}[h!]
	\centering
		\includegraphics[scale=.13]{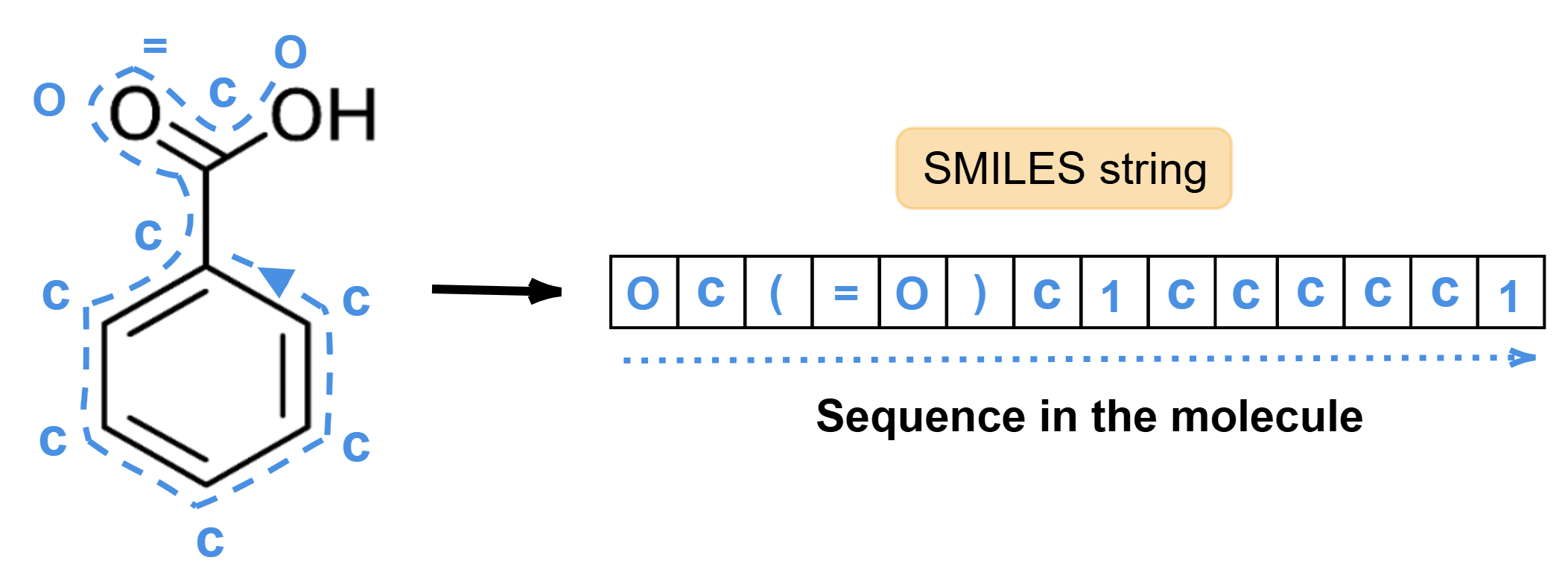}
	\caption{Illustration of the molecular representation of SMILES.}
	\label{fig:Smiles}
 \end{figure}

A SMILES string represents a molecule as a sequential arrangement of alphanumeric symbols that encode its atomic composition and bonding topology \citep{Weininger1988}. Individual atoms are identified by their chemical symbols, while the connectivity between them is delineated by specific characters that denote bond types, branching points, and ring closures. The notation is flexible enough to incorporate stereochemical markers, charges, and functional group patterns, enabling an efficient linear representation of complex structural features, as exemplified in Figure~\ref{fig:Smiles}. This structural encoding is particularly valuable because it establishes a direct link between molecular descriptors and the data-driven models used for viscosity prediction. Therefore, this molecular representation implicitly encodes chemical information that can be passed to a model, indirectly infusing it with domain knowledge.

The dataset employed in this study originates from the work of Pawan Panwar \cite{dataset_viscosity} and was further refined in subsequent research  Panwar et al. \cite{viscosity_dataset_article}. It contains experimental viscosity measurements for 305 hydrocarbon compounds, covering a broad range of molecular architectures and thermophysical conditions. For each compound, the dataset provides the molecular structure in SMILES format, the molecular weight, and dynamic viscosity values measured at five temperatures: 0, 20, 37.77, 60, and 98.89 °C. These temperature points were chosen to capture the intrinsic non-linear dependence of viscosity on thermal conditions and to encompass both low- and high-temperature regimes relevant to petroleum and chemical processing applications.

\begin{figure}[h!]
	\centering
		\includegraphics[scale=.13]{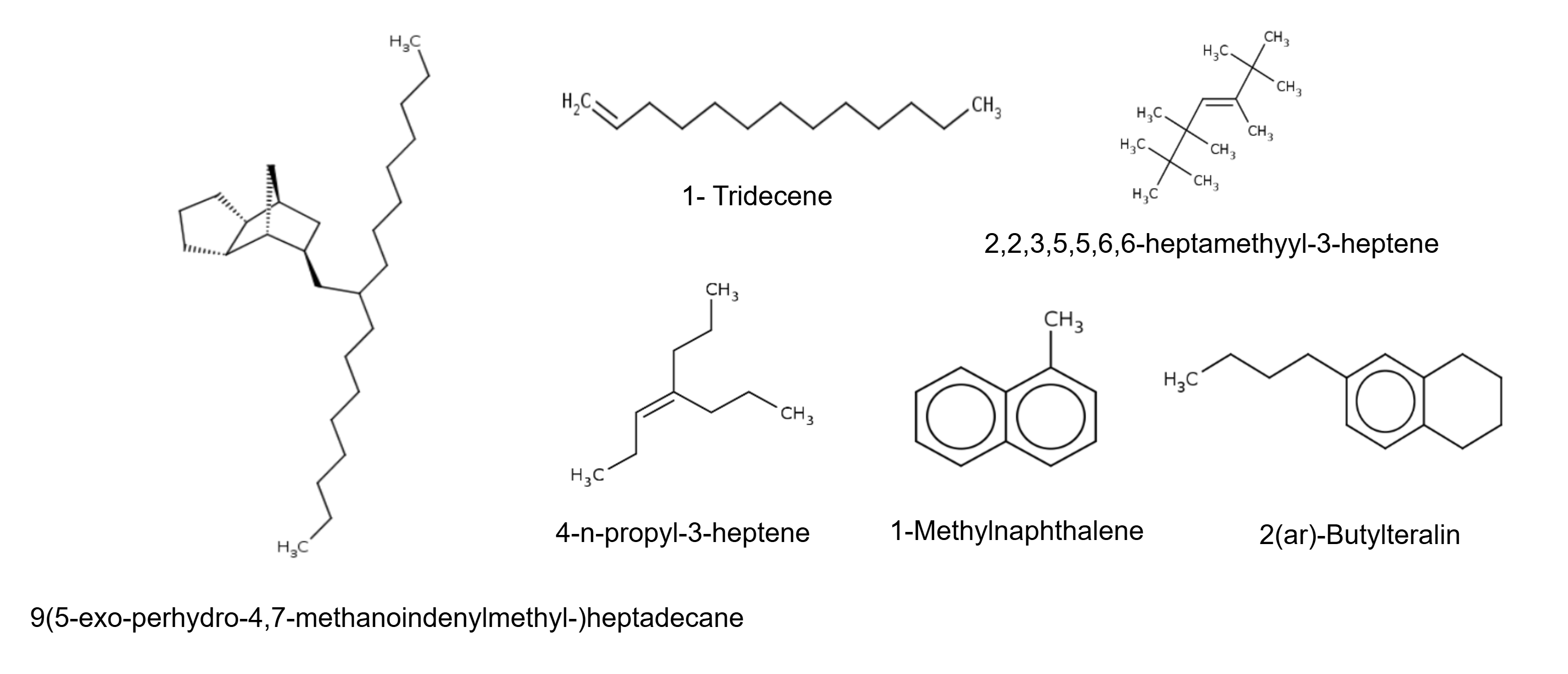}
	\caption{Examples of the hydrocarbon structures that are part of the dataset.}
	\label{fig:molecules}
 \end{figure}

The molecular set spans several hydrocarbon families, including n-paraffins, branched paraffins, 1-olefins, branched olefins, nonfused-ring naphthenes, fused-ring naphthenes, nonfused-ring aromatics, and fused-ring aromatics, as shown in some examples in the Fig. \ref{fig:molecules}. This structural diversity provides a robust foundation for developing machine learning models capable of generalising across a wide spectrum of petroleum molecular configurations.

In addition to the standardisation of molecular representations, the dataset underwent a data treatment process to ensure its quality before modelling. First, records containing missing values or inconsistencies in viscosity measurements or structural information were removed. Then, the curated dataset was organised and split into training, validation, and test subsets, ensuring that the models were evaluated impartially and that their generalisation capability was properly verified.

\subsection{Graph Neural Network and Molecular graph representation} \label{model_tail}
In order to include domain information into the model input, the relation between the molecular structure and the viscosity was obtained through a graph convolutional network (GCN) and a molecular graph representation. This representation is itself a structured encoding of chemical domain knowledge: each molecule is cast as an undirected graph $G = (V, E)$, where the nodes $V$ correspond to atoms and the edges $E$ correspond to covalent bonds between atom pairs. Molecular graphs were generated from SMILES strings using RDKit, so that the chemical identity and bonding rules contained in SMILES are transferred into the graph. Atomic descriptors were assembled into a node feature matrix $X \in \mathbb{R}^{N \times C}$, where $N$ is the number of atoms in the molecule and $C$ is the number of atom-level features. These descriptors are chemically relevant by construction, and thus inject atom-level domain information into the model input. The molecular connectivity was encoded by an adjacency matrix $A \in \mathbb{R}^{N \times N}$, whose elements specify whether two atoms in the molecule are directly bonded. This matrix defines the molecular topology implied by chemistry and determines which atomic features are exchanged during the graph convolution operations. The value $A_{i,j}=1$ if the atom pair $(i,j)$ is bonded, and it is 0 otherwise.

The graph nodes were characterised through a feature vector comprising: \textit{a1}) atom degree, defined as the number of directly bonded neighbouring atoms; \textit{a2}) atomic hybridisation state; \textit{a3}) aromaticity; and \textit{a4}) ring size. Each of these quantities is a standard chemical descriptor, so the node feature matrix $X$ systematically encodes local chemical environments in a numerical form. These features, except \textit{a4}, were encoded using one-hot representations with the default mappings provided by \texttt{PyTorch Geometric 2.6}. Although bond-level descriptors such as bond type and conjugation were initially extracted, they were not included in the final model. This reflects the formulation of the GCN layer adopted in this work, which aggregates information exclusively through node features and the adjacency matrix and therefore does not incorporate edge attributes. In this framework, the adjacency matrix still conveys the chemically defined bonding structure by specifying how atomic features are shared among neighbouring atoms during message propagation.

The molecular graph information, as described above, is mapped to a molecular embedding through GCN. The molecular embedding is a one-dimensional molecular representation with latent features relevant for the predictive task, and it is further used to predict the viscosity as described in Section \ref{model_and_training_details}.

To convert this graph information into numerical embeddings, we employ a GCN including graph convolutional layers following the formulation proposed by Kipf and Welling \cite{kipf2017gcn}. The purpose of a graph convolutional layer is to combine features of each atom with those of its neighboring atoms, so that each atomic representation incorporates information about its local chemical environment. In other words, the convolution operation operationalizes a basic chemical prior: atoms influence, and are influenced by, the atoms to which they are covalently bonded.

The layerwise update rule is given by
\begin{equation}
H^{(l+1)} = \sigma\!\left( \tilde{D}^{-\frac12} \tilde{A} \tilde{D}^{-\frac12} H^{(l)} W^{(l)} \right),
\end{equation}
where $H^{(l)}$ denotes the node feature matrix at layer $l$ with $H^{(0)} = X$. The matrix $\tilde{A} = A + I$ is the adjacency matrix with added self-connections, ensuring that each atom’s original chemical descriptors remain part of its representation while it receives information from its bonded neighbors. The diagonal degree matrix $\tilde{D}$ is defined as $\tilde{D}{ii} = \sum_j \tilde{A}{ij}$. The normalized adjacency operator $\tilde{D}^{-\frac12} \tilde{A} \tilde{D}^{-\frac12}$ performs neighborhood averaging over chemically bonded atoms in a degree-balanced manner, which reflects the idea that an atom’s environment should be aggregated without over-weighting highly connected atoms. The weight matrix $W^{(l)}$ is trainable and projects the aggregated, chemically informed features into a new representation space, while $\sigma(\cdot)$ denotes a nonlinear activation function. In this work, we employed Rectified Linear Unit (ReLU) as the activation function.

Each graph convolutional layer therefore executes two combined operations: (i) aggregation of neighboring atomic features through the normalized adjacency operator, where “neighbors” are defined strictly by covalent bonding in $A$, and (ii) transformation of the aggregated features via a learned linear mapping followed by a nonlinear activation. Stacking multiple layers allows information to propagate across increasingly larger bonded neighborhoods, so the representation evolves from encoding local chemical descriptors (first layers) to encoding larger-scale structural motifs (deeper layers) that are relevant to macroscopic properties such as viscosity.

After the final graph convolutional layer, each atom is associated with an embedding vector that encodes both its original chemical descriptors and the structural context induced by the bonding graph. A global pooling function is then applied over all atoms in a molecule to generate a fixed-length molecular embedding independent of molecular size. This pooled vector is therefore a systematic summary of chemical-based, graph-propagated information, and it serves as the input to the regression network used to predict liquid viscosity.

Once the molecular embedded description is obtained, this is further processed into a feed-forward neural network (FFNN), for computing the viscosity where the model structure is also modified by the inductive biases, so thermodynamic consistency is ensured into the model output.

\subsection{Data-driven prediction and inductive biases} \label{model_head}
In this work, we compare the performance of models incorporating inductive layers with that of a fully data-driven baseline. The motivation for these inductive biases follows common practice in the viscosity literature. Historically, liquid viscosity has been predicted either by purely empirical temperature–viscosity correlations or by semi-empirical forms motivated by thermodynamic arguments. In both cases, the predictive model is typically expressed as a low–order parametric function of temperature, whose coefficients are fitted to experimental data for each pure component. These correlations have remained widely used because they encode, in a compact mathematical form, a temperature dependence that has been repeatedly validated for viscosity.

From a thermodynamic and fluids perspective, such empirical correlations are developed by selecting a functional form that is flexible enough to fit observed temperature trends, while also being smooth and monotonic over the range of interest. Polynomial or rational forms in $T$ (or $1/T$) are especially common because their structure yields stable interpolation and extrapolation across temperature windows where viscosity varies strongly. Importantly for our setting, these equations already embody domain understanding about how viscosity responds to temperature, i.e., they provide a tested functional scaffold for viscosity prediction.

However, classical use of these correlations is limited by their component-specific nature. The parameters of each correlation are identified for a given liquid from dedicated experimental datasets, and are only valid under the conditions for which they were calibrated. When a new component is considered, a new set of parameters must be re-fitted, which restricts transferability and hinders large-scale screening. Thus, while the functional forms are broadly useful, their conventional parameterization is not.

Our approach is to retain the usefulness of these established functional forms (i.e., the domain-tested mapping from temperature to viscosity) while replacing the component-wise parameter fitting with a learned, chemistry-informed parameter prediction. After including chemical and structural domain information into the input through the molecular graph and GCN embedding, we further include domain information into the output by constraining the model to predict parameters of viscosity correlations that are known to work well for this property.

Two inductive biases were employed, namely the \textit{Andrade inductive bias} and the \textit{empirical inductive bias}. The Andrade inductive bias employs a variant of the Andrade equation for describing viscosity, as reported by \cite{Gutmann_and_Simmons_1952}, which relates viscosity to temperature through a rational dependence on $T$. Specifically, the formulation employs 3 fitting parameters and is described as:

\begin{equation}
\label{andrade_equation}
\log_{10}(\eta) = A + \frac{B}{T + C}.
\end{equation}
Here, $\eta$ is the viscosity, $A$, $B$, and $C$ are liquid-specific parameters, and $T$ is the temperature in Kelvin. This equation has long been used to capture the steep, nonlinear temperature sensitivity of viscosity in many liquids, and its constrained form provides a thermodynamically consistent and smooth temperature trend.

The other alternative we propose for the inductive bias, the empirical one, employs an empirical equation often used in the petrochemical industry to regress liquid viscosity data \cite{Cocker_2007}. This correlation augments inverse–temperature dependence with low-order polynomial terms in $T$, allowing additional curvature when required by experimental trends.

\begin{equation}
\label{empirical_equation}
\log_{10}(\eta) = A + \frac{B}{T} + C\cdot T + D\cdot T^2.
\end{equation}
Here, $\eta$ is the viscosity, $A$, $B$, $C$ and $D$ are liquid-specific parameters, and $T$ is the temperature in Kelvin. While still empirical, this polynomial structure is a standard, domain-validated way to represent viscosity–temperature behavior over broad ranges.

We propose an ANN architecture that incorporates inductive-bias modifications in its final layer. Specifically, the standard output head is replaced by a viscosity correlation that acts as a domain-informed activation function. After the molecular graph is encoded into an embedding (via GCNs and pooling), the network produces intermediate outputs that are passed through the Andrade form \eqref{andrade_equation} or the empirical form \eqref{empirical_equation}. The output of this inductive layer is therefore the predicted $\log_{10}(\eta)$, computed through a functional form known to represent viscosity–temperature behavior.

This architectural change shapes training in a systematic way. During the forward pass, every prediction must flow through the chosen correlation, so the model is constrained to generate viscosity values that follow the established temperature dependence encoded by that equation. During backpropagation, the prediction error is differentiated through the inductive layer and its analytic temperature–viscosity mapping. Consequently, gradients propagated to earlier layers are modulated by this domain-informed transformation, meaning that parameter updates are guided jointly by data fit and the imposed functional structure. In contrast, the baseline data-driven model feeds the molecular embedding and temperature into a standard neural head that outputs $\log_{10}(\eta)$, so its gradients are not shaped by any viscosity-specific mathematical functional forms.

\subsection{Model and training details}
\label{model_and_training_details}
All the models employed in this work employed the same network for converting the molecular graph representation into molecular embedding. In this paper, we employed a series of 3 graph convolutional layers, preceded by a graph normalisation layer using GraphNorm layer (\cite{cai2021GraphNorm}). The resulting vector was pooled using column-wise summation. 
To increase the convergence stability, the obtained embedding values were normalised using batch normalisation (\cite{ioffe2015BatchNorm}).

Following the computation of the molecular embedding, a FFNN is used to predict the viscosity or the parameter used in the inductive bias. In the data-driven model, the FFNN takes as input the experimental temperature along with the molecular embedding, and having as input only one node to predict directly $log_{10}(\eta)$. On the other hand, the FFNN of the models, including inductive bias, takes as input the computed molecular embedding, and has a number of output nodes equal to the number of parameters to predict in the inductive bias (i.e., 3 for the Andrade and 4 for the empirical inductive biases). The number of hidden layers and the number of nodes contained in them are the same for all the networks. In all the networks, the hidden nodes were activated using a ReLU function, while the output node was activated with a linear function.

The model hyperparameters are included in Table \ref{tab:model_hyperparameters}. These were obtained through a manual trial-and-error approach and guaranteed stability and accurate predictions on the validation set.

\begin{table}[H]
\renewcommand{\arraystretch}{1.3}
\caption{Selected hyperparameter values for the GCN+MLP models.\label{tab:model_hyperparameters}}
\centering
\small
\begin{tabularx}{0.6\linewidth}{@{}%
    >{\raggedright\arraybackslash}p{1.8cm}  
    >{\raggedright\arraybackslash}p{4cm}  
    >{\centering\arraybackslash}p{2cm}    
@{}}
    \toprule
    \textbf{Model} & \textbf{Hyperparameter} & \textbf{Selected value} \\
    \midrule
    GCN & Hidden channels      & 128 \\
        & Number of layers     & 3   \\
        & Dropout              & 0 \\
        & Output channels      & 128 \\
    \midrule
    MLP  & Hidden channels      & 50  \\
         & Number of layers     & 4   \\
         & Dropout              & 0.0 \\
    \bottomrule
\end{tabularx}
\end{table}

All models were implemented using the PyTorch Geometric framework \cite{fey2025pyg20scalablelearning}, employing the graph convolutional network (GCN) architecture introduced by Kipf and Welling \cite{kipf2017gcn}. Training was conducted using the PyTorch Ignite framework \cite{pytorch-ignite}. Optimization was performed with the Adam optimizer \cite{kingma2017adam}, using a learning rate of $1\cdot10^{-2}$ and a batch size of 32. The Huber loss was used during training, while model performance was assessed on both the validation and test sets using the mean squared error (MSE), the mean absolute error (MAE), and root mean squared error (RMSE).
Models were trained for 500 epochs, and both the model predictions and the experimental data were normalized before evaluation in the loss function. During training, the MSE on the validation set was continuously monitored and used as the checkpointing criterion: whenever the validation MSE improved compared to the previous best value, the model weights were saved. The final selected model corresponds to the checkpoint achieving the lowest validation MSE across all training epochs. Further details about model training are included in Algorithm \ref{alg:gnn_visc}.
Due to the presence of missing experimental viscosity values at certain temperature points for some molecules, both the training loss and evaluation metrics were computed using protected formulations. Specifically, predictions corresponding to missing experimental measurements were excluded from loss and metric calculations, ensuring that only valid data points contributed to the optimization and performance assessment. Furthermore, to increase the training stability, the temperature has been divided by 273.15 when employing it in the inductive biases.
\begin{algorithm}[H]
\caption{GNN-based Viscosity Prediction with Detailed Embedding}
\label{alg:gnn_visc}
\KwIn{Set of molecules $\{M_i\}_{i=1}^N$, viscosities $\{y_i\}_{i=1}^N$, training parameters}
\KwOut{Trained model capable of predicting viscosity}

\tcp{Step 1: Preprocess molecules and build graphs}
\For{$i \gets 1$ \KwTo $N$}{
  $G_i \gets \text{ConvertSMILESToGraph}(M_i)$\;
}
\tcp{Step 2: Initialize model parameters}
$\theta \gets \text{InitializeParameters}()$\;

\tcp{Step 3: Training loop}
\ForEach{epoch $e \in \{1, \ldots, E\}$}{
  \For{$i \gets 1$ \KwTo $N$}{
    \tcp{Step 3.1: GNN Embedding Calculation}
    \tcp{For each node $v$ in $G_i$, update:}
    \tcp{$h_v^{(l)} = \sigma\Big(\sum_{u \in \mathcal{N}(v)} W^{(l)}\,h_u^{(l-1)} + b^{(l)}\Big)$ for $l=1,\dots,L$}
    \tcp{Aggregate node features to form graph-level embedding:}
    \tcp{$H_i = \text{POOL}(\{h_v^{(L)} : v \in G_i\})$, i.e.,}
    \[
      H_i = f(G_i; \theta) = \mathrm{POOL}\Big(\{h_v^{(L)} : v \in G_i\}\Big)
    \]
    \tcp{(Here, $\sigma$ is an activation function, $\mathcal{N}(v)$ is the set of neighbors of node $v$, and $W^{(l)}$, $b^{(l)}$ are learnable parameters.)}
    $H_i \gets \text{GNN\_Embedding}(G_i; \theta)$\;
    
    $Z_i \gets \text{FullyConnected}(H_i; \theta)$\;
    $(A_i, B_i, C_i) \gets \text{OutputLayer}(Z_i; \theta)$\;
    $y_i^\text{pred} \gets \text{PredictViscosity}(A_i, B_i, C_i)$\;
    $\text{Loss}_i \gets \text{LossFunction}(y_i^\text{pred}, y_i)$\;
  }
  $\text{Loss} \gets \frac{1}{N}\sum_{i=1}^{N} \text{Loss}_i$\;
  \tcp{Backpropagation and parameter update}
  $\theta \gets \theta - \alpha \nabla_\theta \text{Loss}$\;
}
\end{algorithm}

\section{Results and Discussions}
The dataset used in this study comprises 305 pure hydrocarbon molecules spanning a broad range of structural families, including linear and branched paraffins, olefinic species, nonfused and fused naphthenes, and aromatic systems of varying degrees of complexity. This molecular diversity is reflected in the wide distribution of molecular sizes, with molar masses ranging from approximately $110.20 \,\text{g mol}^{-1}$ to $703.30 \,\text{g mol}^{-1}$, ensuring representation of both relatively small hydrocarbons and substantially larger species within the typical range of petroleum-related compounds.

The dataset covers a similarly extensive thermophysical space. The dynamic viscosity measurements span nearly five orders of magnitude, from about $0.29 \,\text{cP}$ for low-viscosity liquids to values close to $2.00 \times 10^{4} \,\text{cP}$ for highly viscous fluids. Densities fall within the expected interval for liquid hydrocarbons, ranging roughly from $0.67$ to $1.12 \,\text{g cm}^{-3}$. Measurements were performed at temperatures between approximately $0$ and $135\,^{\circ}\text{C}$, ensuring that the models are trained across a broad thermal range where the temperature sensitivity of liquid viscosity is particularly pronounced.

To ensure a robust assessment of the models' generalization capabilities, the dataset was randomly partitioned into three independent subsets: $80\%$ for training, $10\%$ for validation, and $10\%$ for final testing. The data were structured molecule-wise, and each point contained both molecular information and experimental viscosity values at different temperatures. The split was done on the molecules. This strategy prevents hyperparameter selection from being influenced by data used during model fitting and guarantees that the final evaluation is conducted exclusively on molecular structures not seen during model development.

\begin{figure}[h!]
	\centering
		\includegraphics[scale=.6]{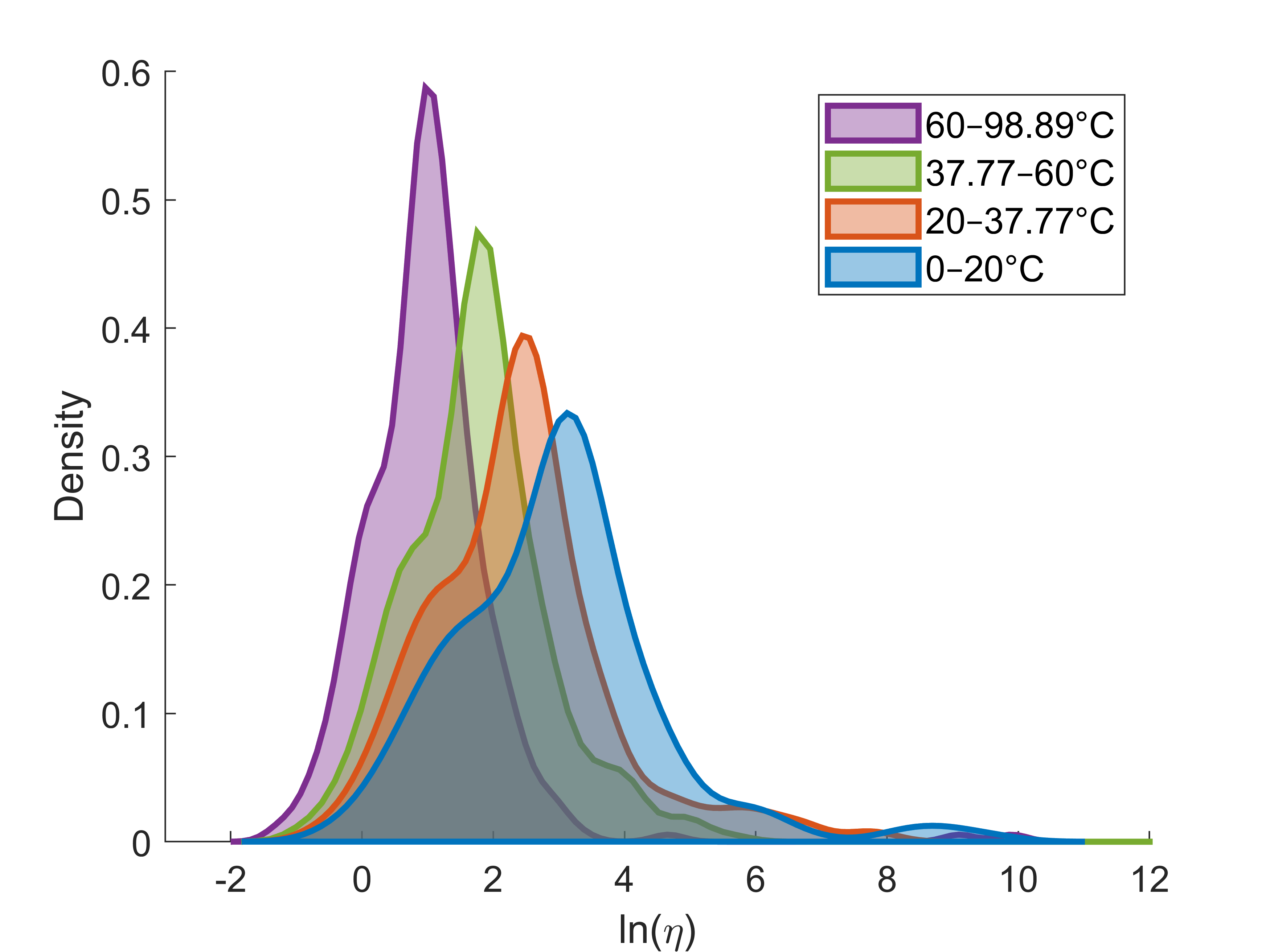}
	\caption{Distribution of logarithmic viscosity values in the curated hydrocarbon dataset across the experimental temperature range.}
	\label{Histogram_dataset}
 \end{figure}

Fig. ~\ref{Histogram_dataset} illustrates the distribution of logarithmic viscosities across all temperatures included in the study. A substantial spread is observed, reflecting both the structural heterogeneity of the molecules and the wide thermodynamic range they cover. This variability highlights the challenge for predictive modeling, as the model must account for significant differences across chemical families while also capturing the nonlinear relationship between viscosity and temperature. Such characteristics reinforce the need for approaches capable of representing global trends and complex thermophysical behavior, particularly in the high-viscosity regime, where predictive accuracy is often most relevant.

Based on the previously described curated dataset, a series of neural network models were trained using the architectural configurations summarized in Table~\ref{tab:model_hyperparameters}. The hyperparameter values listed were selected after a targeted sensitivity analysis that systematically evaluated the influence of key architectural choices, including hidden dimensionality, depth, dropout rate, and regularization. This procedure ensured that the final configurations provided an appropriate structure for the models.

Three modeling strategies were investigated. The first was a fully data-driven baseline model in which the network directly maps the molecular graph representation and temperature to the corresponding logarithmic viscosity values. In addition to this baseline, two hybrid approaches were developed within the ExPUFFIN framework. The first integrates the Andrade-type formulation as an inductive bias, imposing the functional structure expressed in Equation~\ref{andrade_equation}. The second employs an extended empirical correlation represented in Equation~ \ref{empirical_equation}, which allows for a more flexible temperature dependence.

\begin{figure}[h!]
    \centering

    \begin{subfigure}[b]{0.32\textwidth}
        \centering
        \includegraphics[width=\linewidth]{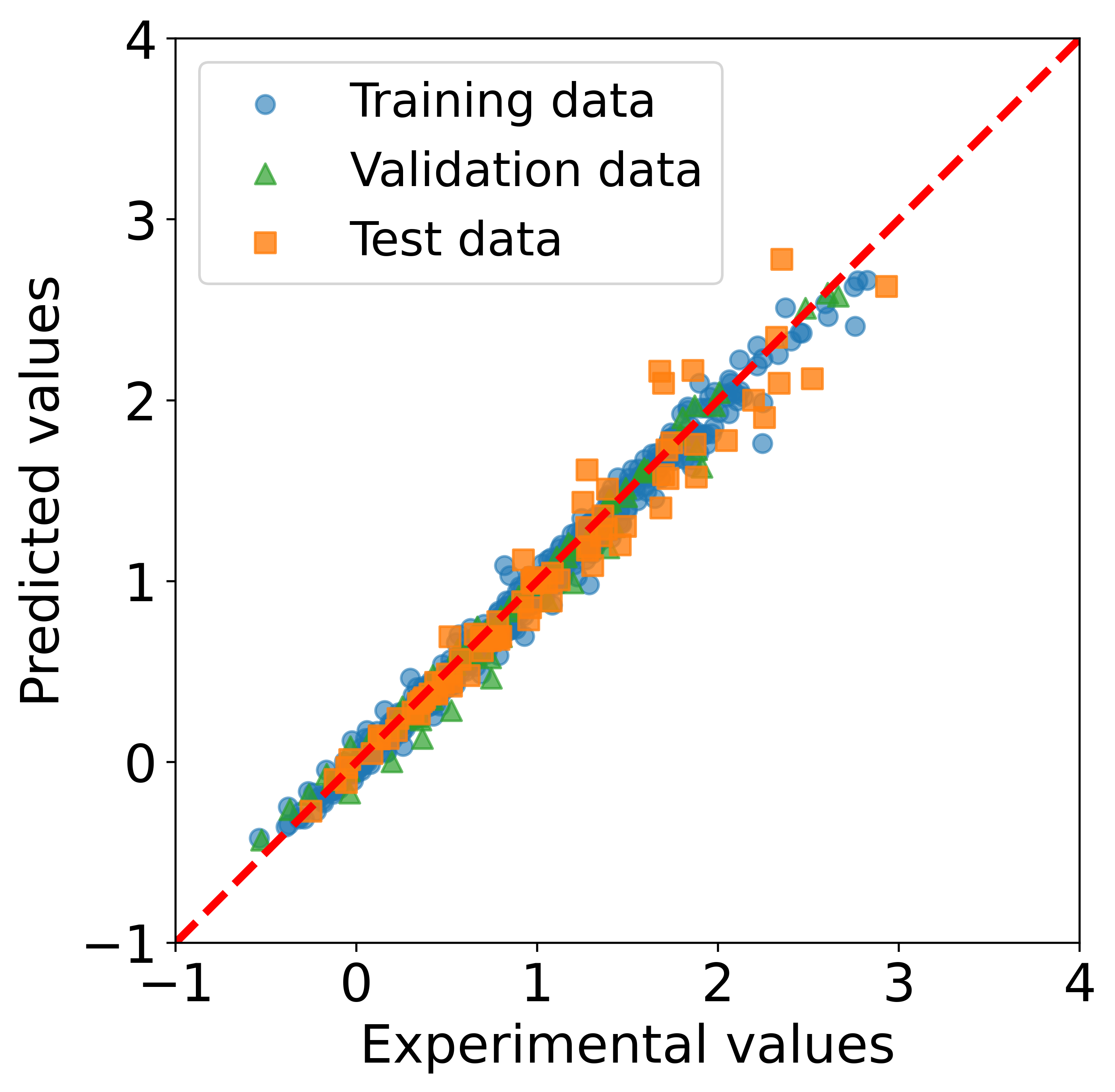}
        \caption{Baseline model.}
        \label{fig:Baseline_parity}
    \end{subfigure}
    \hspace{0.015\textwidth}
    \begin{subfigure}[b]{0.32\textwidth}
        \centering
        \includegraphics[width=\linewidth]{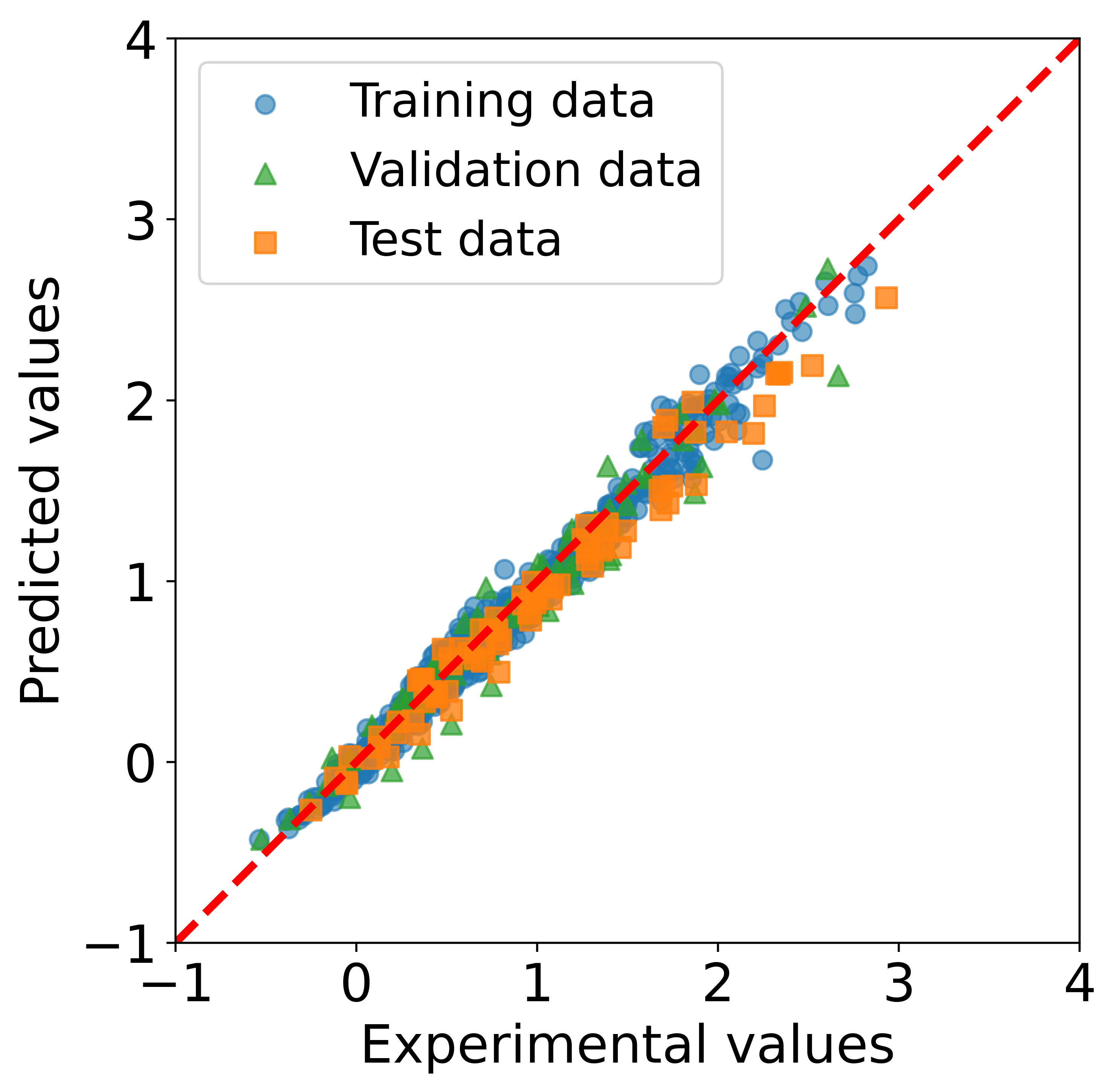}
        \caption{ExPUFFIN Andrade model}
        \label{fig:Andrade_parity}
    \end{subfigure}
        \begin{subfigure}[b]{0.32\textwidth}
        \centering
        \includegraphics[width=\linewidth]{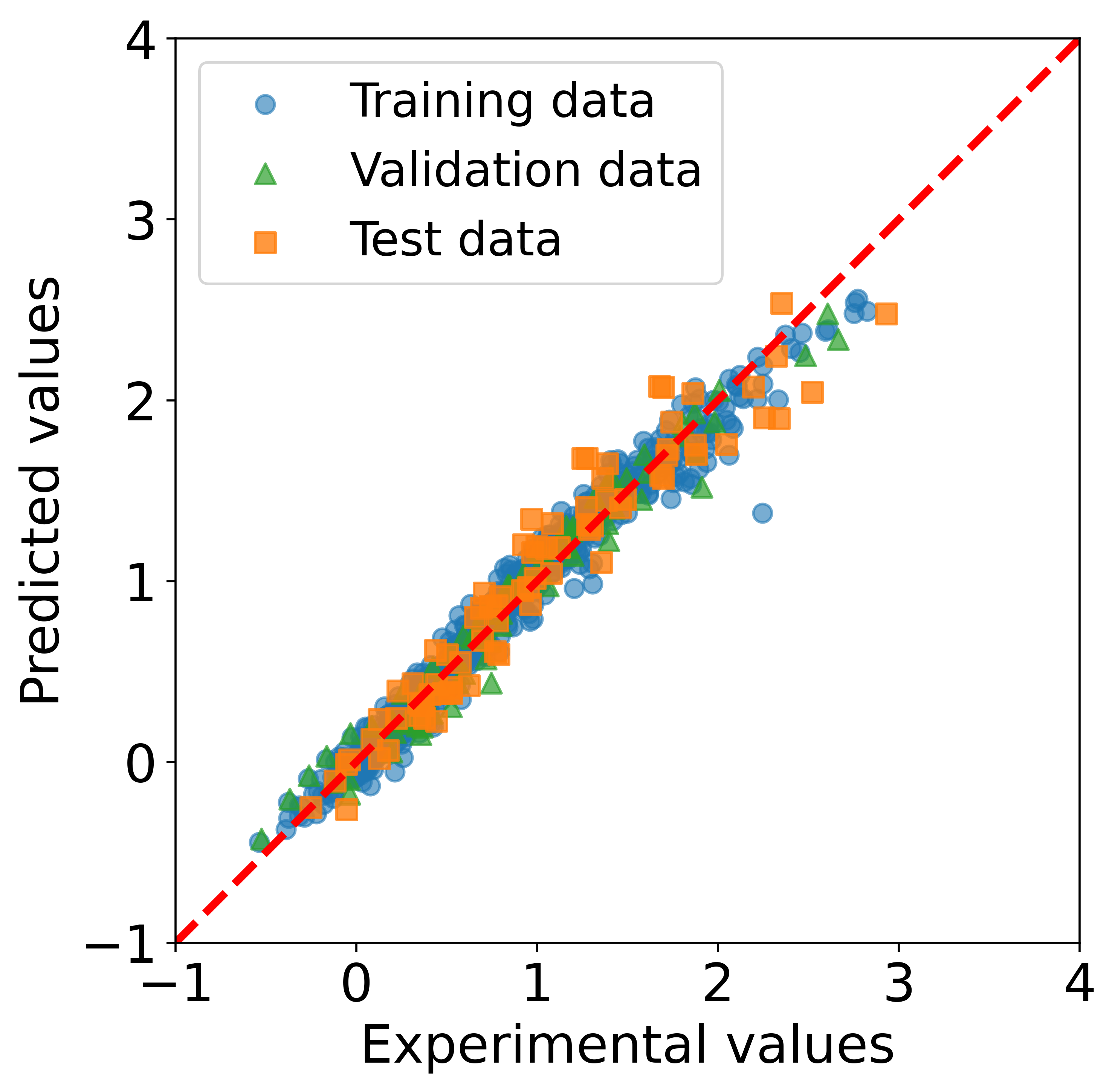}
        \caption{ExPUFFIN empirical model}
        \label{fig:empirical_parity}
    \end{subfigure}

    \caption{Parity plots comparing predicted and experimental viscosities for the baseline, ExPUFFIN, and ExPUFFIN empirical models.}
    \label{fig:Parity_plots}
\end{figure}

The resulting predictions are illustrated in the parity plots shown in Fig. ~\ref{fig:Parity_plots}, where the agreement between predicted and experimental logarithmic viscosities can be directly examined for the baseline model, the ExPUFFIN Andrade model, and the ExPUFFIN empirical model.

Table~\ref{tbl:MAE_MSE_Test_Dataset} summarizes the predictive accuracy of the three architectures on the independent test set. All models achieve low errors (RMSE $\approx 0.15$–$0.19$ in $\log_{10}\eta$) and the parity plots in Fig.~5 show clustering around the ideal $y=x$ line across training, validation, and test partitions. In a standard purely data-driven modelling assessment, these results would lead to the conclusion that the baseline GNN already provides a satisfactory viscosity predictor for the petroleum dataset, since its test errors are comparable to those of the inductive variants.

However, this conclusion must be interpreted in light of what each model is allowed to learn. The baseline model is unconstrained in how it maps temperature and molecular structure to viscosity; it can fit the observed temperature–viscosity trends in the dataset, but nothing in its architecture enforces that these trends remain thermodynamically plausible outside the training domain. Consequently, the baseline model should be regarded as reliable mainly for careful interpolation within the temperature and composition ranges represented in the data. When extrapolating to temperatures not covered by the dataset—or to novel molecules whose viscosity–temperature behavior differs from those seen during training—the baseline model may produce predictions that are numerically reasonable but physically inconsistent, because it has no built-in mechanism to preserve the known functional dependence of viscosity on temperature.

By contrast, both ExPUFFIN variants embed established viscosity–temperature correlations directly in the output layer. While their test errors are of the same order as the baseline, their predictions are generated through a domain-validated functional form at every forward pass. This means that the model is not only fitting the dataset, but doing so under a structural constraint that is known to generalize well for viscosity. The ExPUFFIN Andrade model yields the lowest RMSE and MSE, indicating that the Andrade form provides the most suitable inductive scaffold for this dataset. The ExPUFFIN empirical model performs slightly worse in absolute error, but still maintains the same overall parity behavior, indicating that enforcing a polynomial-type temperature dependence remains compatible with the learned molecular effects.

\begin{table}[h!]
\renewcommand{\arraystretch}{1.4}
\centering
\caption{Final RMSE, MAE, and MSE on the test dataset for the models.}
\label{tbl:MAE_MSE_Test_Dataset}
\begin{tabular}{lccccccccc}
    \toprule
    \multirow{2}{*}{\textbf{Dataset}} 
    & \multicolumn{3}{c}{\textbf{Baseline model}} 
    & \multicolumn{3}{c}{\textbf{ExPUFFIN Andrade model}} 
    & \multicolumn{3}{c}{\textbf{ExPUFFIN empirical model}} \\
    \cmidrule(lr){2-4} 
    \cmidrule(lr){5-7}
    \cmidrule(lr){8-10}
    & \textbf{RMSE} & \textbf{MAE} & \textbf{MSE} 
    & \textbf{RMSE} & \textbf{MAE} & \textbf{MSE}
    & \textbf{RMSE} & \textbf{MAE} & \textbf{MSE} \\
    \midrule
    Petroleum data 
    & 0.1610 & 0.1124 & 0.02592
    & 0.1544 & 0.1243 & 0.02384
    & 0.1866 & 0.1425 & 0.03481 \\
    \bottomrule
\end{tabular}
\end{table}

\begin{figure}[h!]
    \centering

    \begin{subfigure}[b]{0.32\textwidth}
        \centering
        \includegraphics[width=1.02\linewidth]{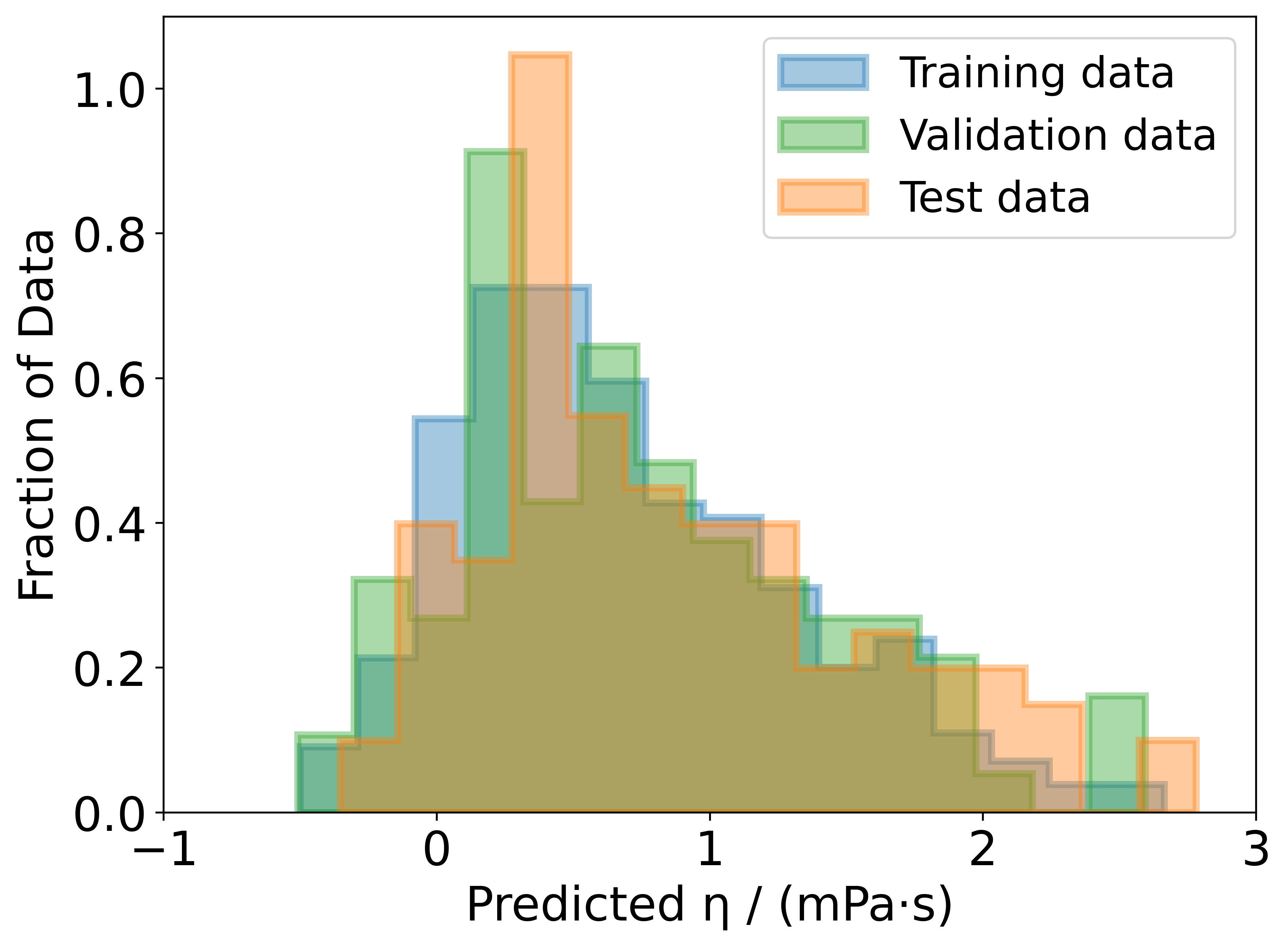}
        \caption{Baseline model.}
        \label{fig:Hist_baseline}
    \end{subfigure}
    \hspace{0.015\textwidth}
    \begin{subfigure}[b]{0.32\textwidth}
        \centering
        \includegraphics[width=1.02\linewidth]{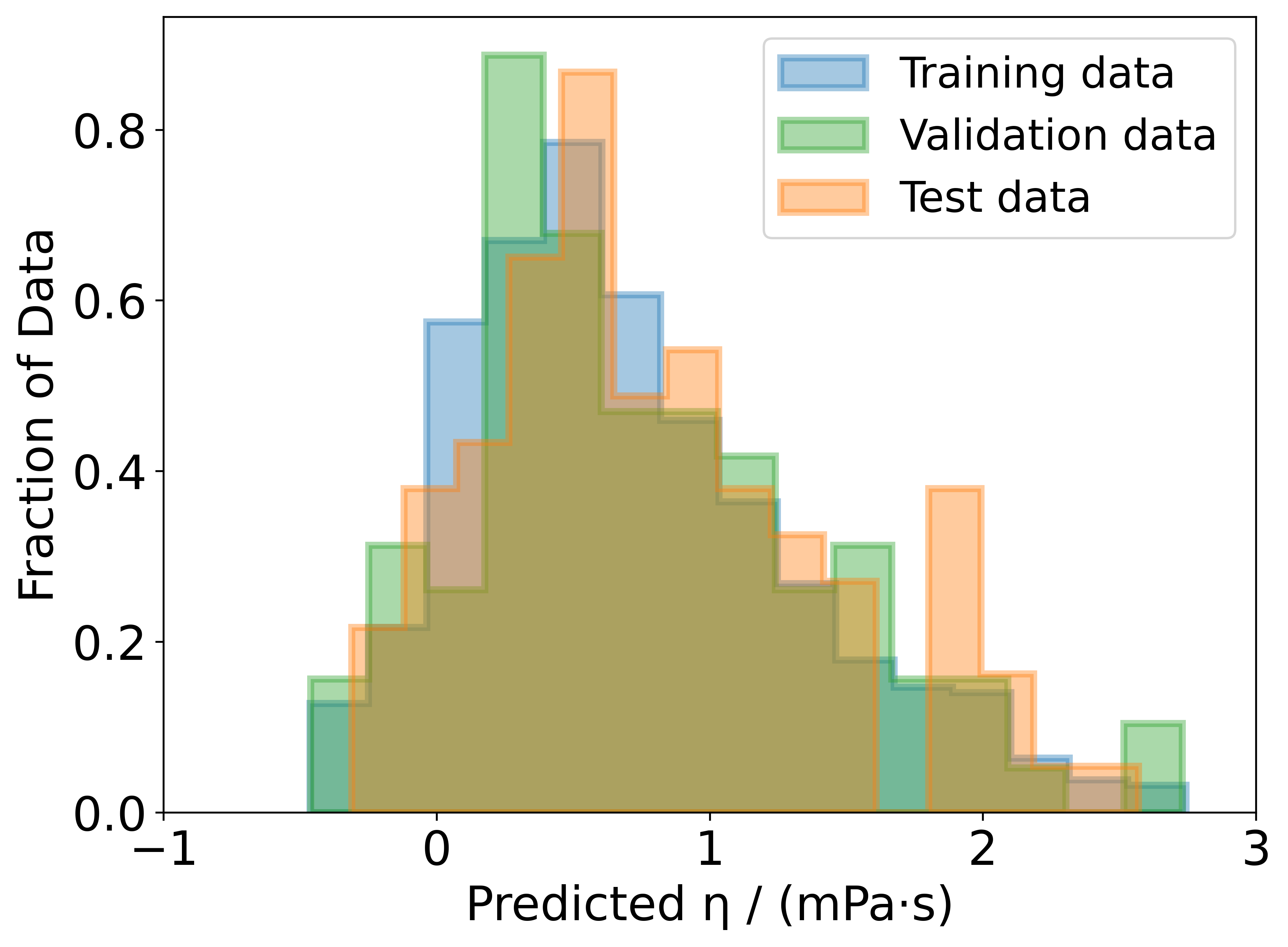}
        \caption{ExPUFFIN Andrade model.}
        \label{fig:Hist_Andrade}
    \end{subfigure}
    \begin{subfigure}[b]{0.32\textwidth}
        \centering
        \includegraphics[width=1.02\linewidth]{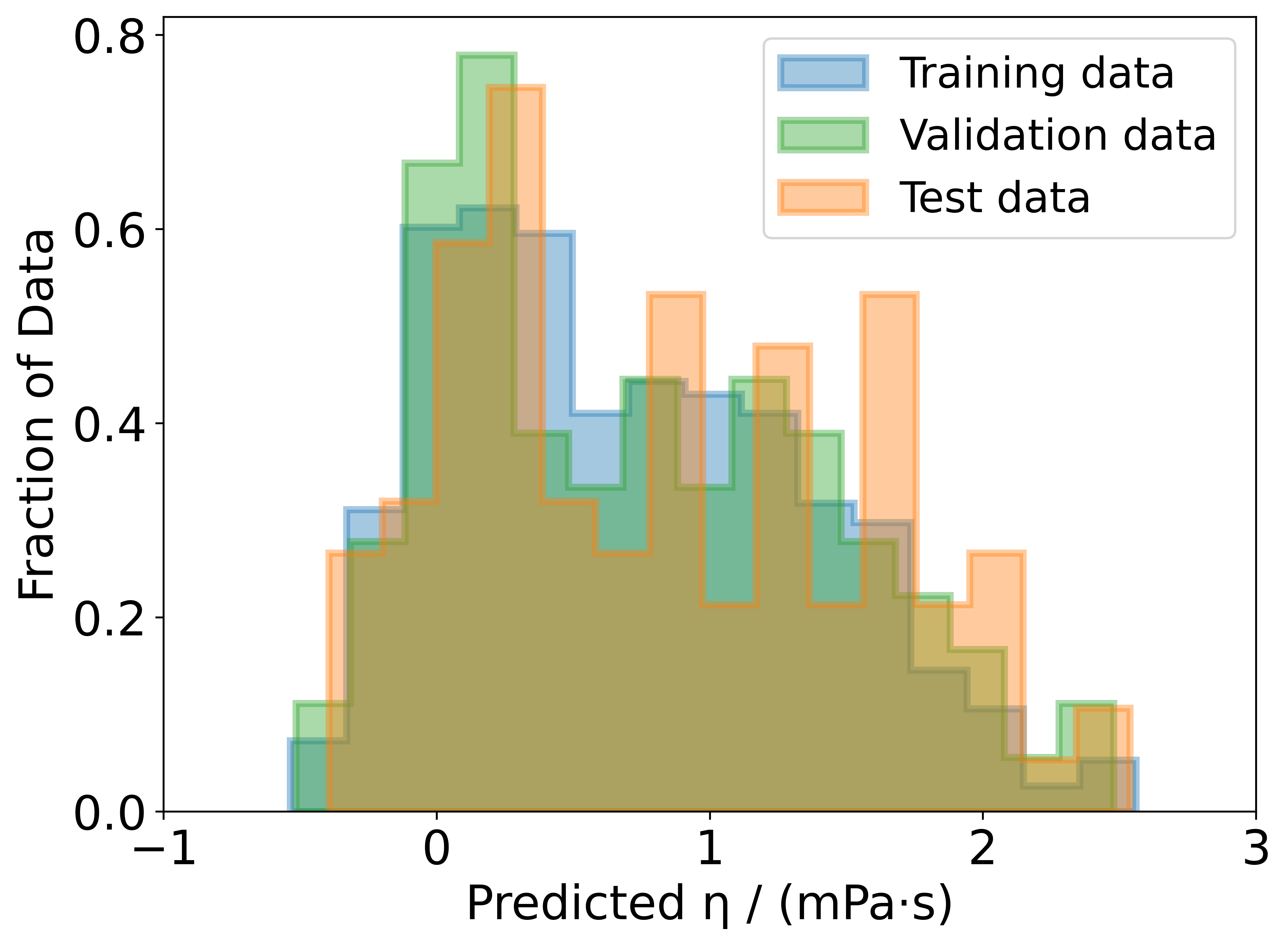}
        \caption{ExPUFFIN empirical model.}
        \label{fig:Hist_empirical}
    \end{subfigure}
    \caption{Distribution of predicted viscosity values across the training, validation, and test datasets.}
    \label{fig:Histogram_predictions}
\end{figure}

In addition to the parity analysis, we examined the distributions of predicted viscosities across the training, validation, and test partitions (Fig.~\ref{fig:Histogram_predictions}). This comparison is important because a model can achieve low aggregate error while still distorting the viscosity range, for example, by systematically compressing high-viscosity values or over-spreading low ones. In that case, good performance metrics may hide weak generalization.

Across all three architectures, the predicted distributions for training, validation, and test data largely overlap and span the same viscosity interval. This indicates that none of the models is suffering from an obvious distribution shift between partitions, and that the learning process did not lead to a biased output range. In particular, the test-set histogram follows the same multimodal structure observed in training and validation, suggesting that the models generalize over the viscosity spectrum present in the dataset rather than only matching its mean.

Subtle differences between models are still visible. The baseline predictions show a slightly sharper concentration around the central viscosity region, consistent with a purely data-driven model that tends to interpolate toward the densest part of the training distribution. The ExPUFFIN Andrade model preserves the overall shape while maintaining a comparable spread into the higher-viscosity tail, indicating that the inductive output layer does not restrict the model’s ability to represent the full range of viscosities. The ExPUFFIN empirical model exhibits a marginally broader and noisier distribution, especially in the mid-to-high region, consistent with its slightly higher test error.

Overall, Fig.~\ref{fig:Histogram_predictions} supports the conclusion drawn from Table~\ref{tbl:MAE_MSE_Test_Dataset} and Fig.~5: all models learn a stable mapping over the viscosity range covered by the data.

\begin{figure}[h!]
    \centering

    \begin{subfigure}[b]{0.32\textwidth}
        \centering
        \includegraphics[width=\linewidth]{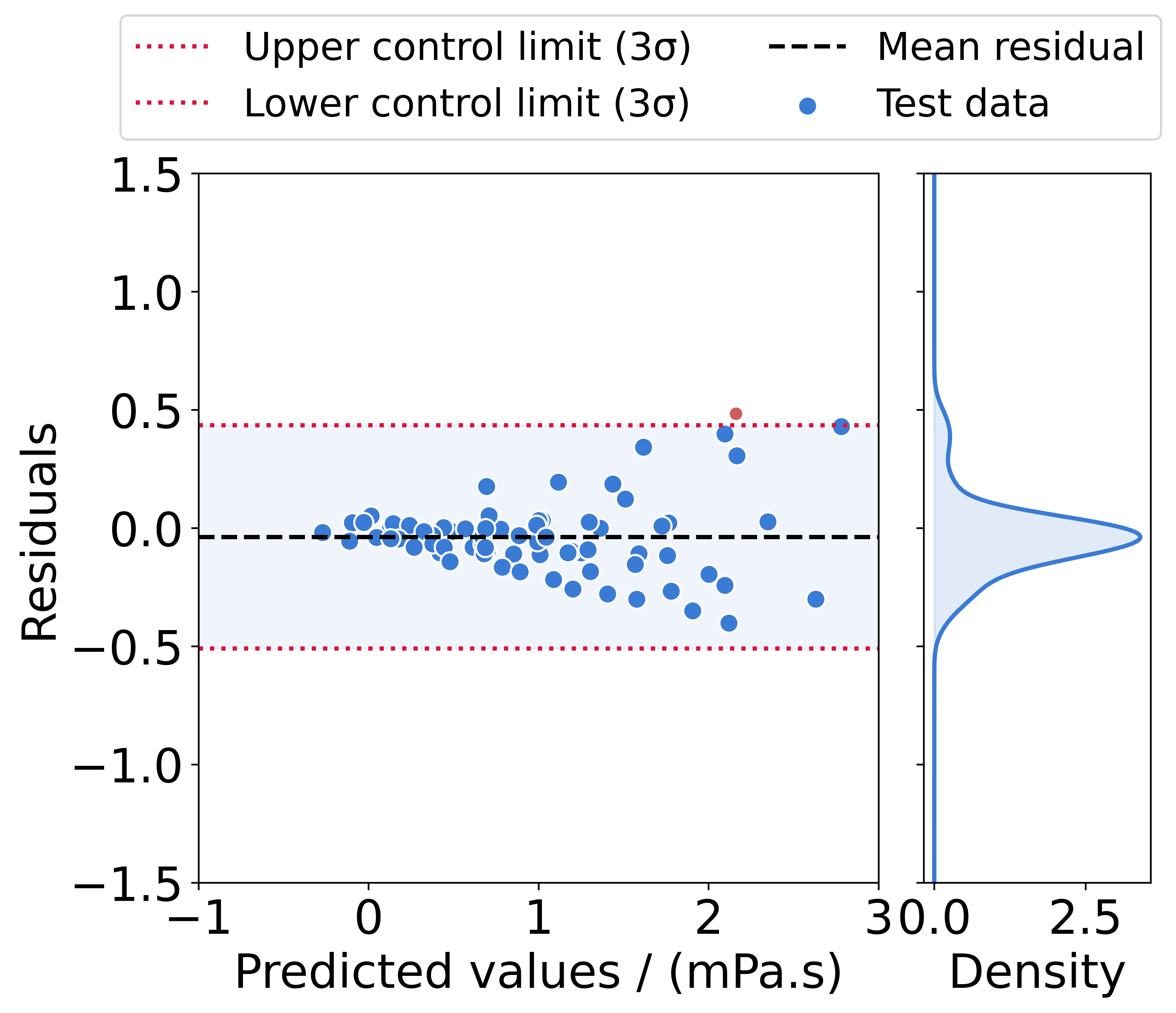}
        \caption{Baseline model.}
        \label{fig:baseline_residuals}
    \end{subfigure}
    \hspace{0.015\textwidth}
    \begin{subfigure}[b]{0.32\textwidth}
        \centering
        \includegraphics[width=\linewidth]{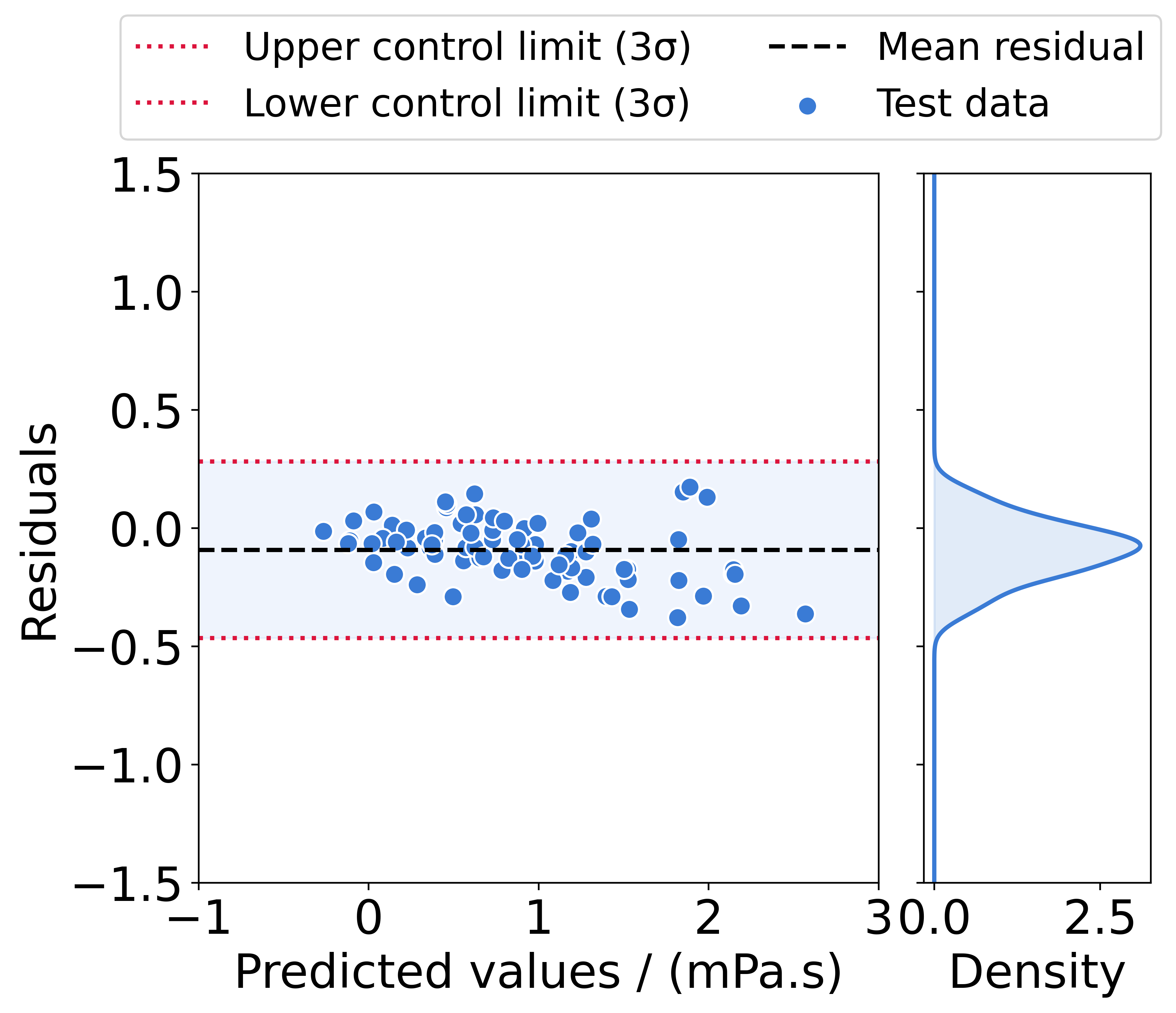}
        \caption{ExPUFFIN Andrade model}
        \label{fig:Andrade_residuals}
    \end{subfigure}
        \begin{subfigure}[b]{0.32\textwidth}
        \centering
        \includegraphics[width=\linewidth]{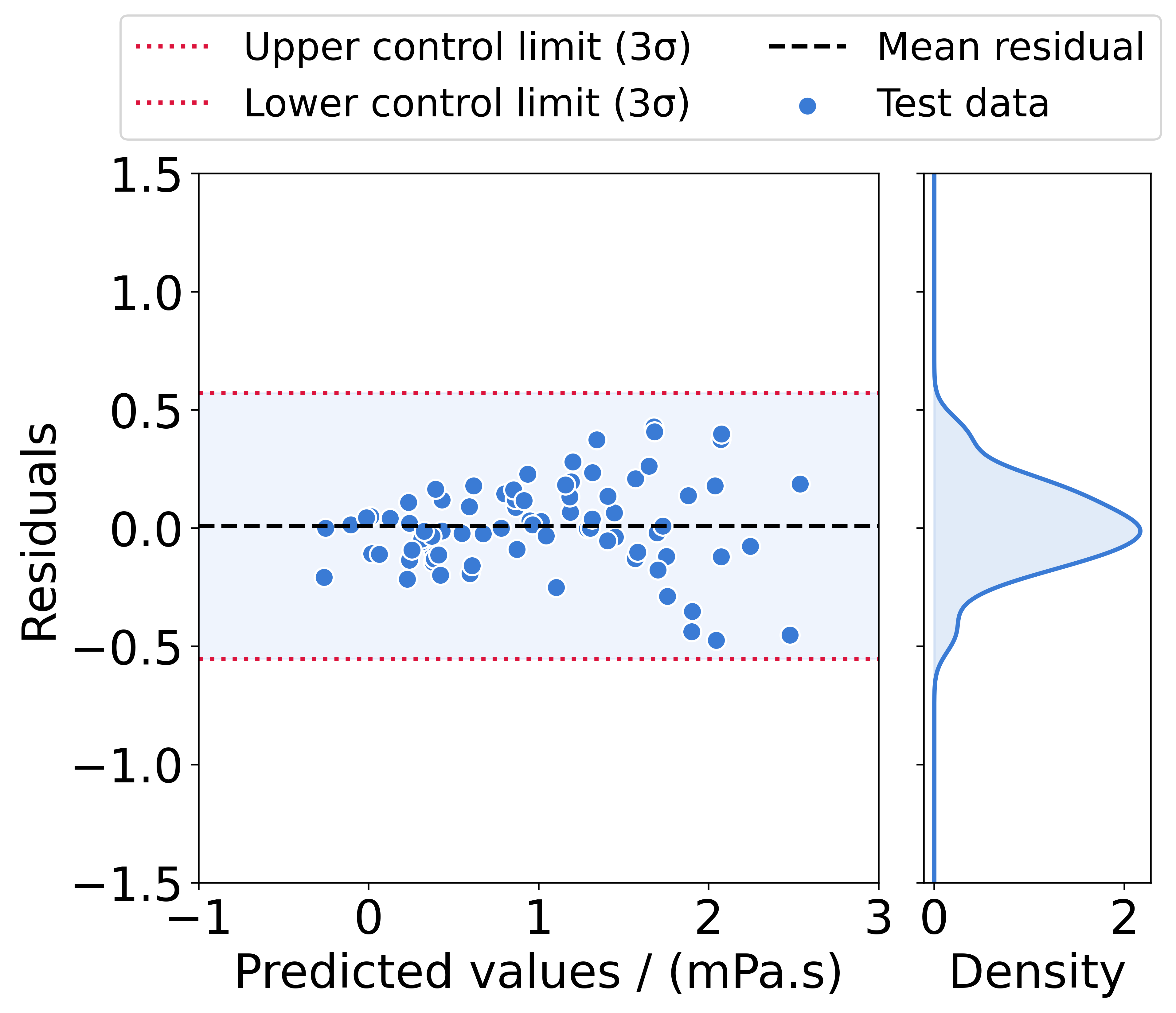}
        \caption{ExPUFFIN empirical model}
        \label{fig:empirical_residuals}
    \end{subfigure}

    \caption{Residual distributions for the baseline, ExPUFFIN Andrade, and ExPUFFIN empirical models.}
    \label{fig:Residuals}
\end{figure}

Residual analysis complements the parity plots by revealing how prediction errors are distributed across the viscosity range and whether any systematic bias remains. Figure~\ref{fig:baseline_residuals}–\ref{fig:empirical_residuals} report residuals on the test set as a function of predicted viscosity, together with their marginal densities and $\pm 3\sigma$ control bands. For all models, residuals are centered close to zero and remain largely within the control limits, confirming the good overall fit already suggested by Table~\ref{tbl:MAE_MSE_Test_Dataset}.

Despite this shared accuracy, the spread of residuals differs across architectures. The baseline model (Fig.~\ref{fig:baseline_residuals}) exhibits a visibly wider scatter, especially as predicted viscosity increases, indicating higher variability in error across the range. The accompanying density is correspondingly broader, meaning that while the model is unbiased on average, it achieves this with less consistent point-wise accuracy.

A tighter and more homogeneous residual pattern is observed for the ExPUFFIN Andrade model (Fig.~\ref{fig:Andrade_residuals}). Residuals cluster closer to zero throughout the viscosity range and their density is narrower, indicating more stable predictive behavior. This supports the role of the Andrade inductive layer: by enforcing a domain-validated temperature–viscosity functional structure during every forward and backward pass, the model learns corrections that are not only accurate but also less dispersed.

The ExPUFFIN empirical model (Fig.~\ref{fig:empirical_residuals}) retains residuals centered near zero, but with a slightly larger dispersion than the Andrade variant, consistent with its higher RMSE and MAE. Nevertheless, the residual cloud remains well-behaved (no obvious drift with viscosity and no heavy tails), indicating that this inductive form still provides a meaningful structural constraint even if it is not as well matched to the present dataset as the Andrade form.

Overall, the residual plots reinforce two points. First, all three models deliver accurate in-range predictions with no strong systematic over- or under-estimation. Second, the inductive-bias architectures—especially ExPUFFIN Andrade—reduce error dispersion, suggesting improved robustness beyond what aggregate metrics alone capture.

In the next two sections, we explicitly investigate this reliability question, analysing the extent to which the ExPUFFIN architecture improves robustness—particularly for interpolation and extrapolation in temperature—and clarifying where the inductive-bias formulation provides measurable benefits over the baseline.

\subsection{Interpolation analysis}

\begin{figure}[h!]
    \centering

    \begin{subfigure}[b]{0.47\textwidth}
        \centering
        \includegraphics[width=1.0\linewidth]{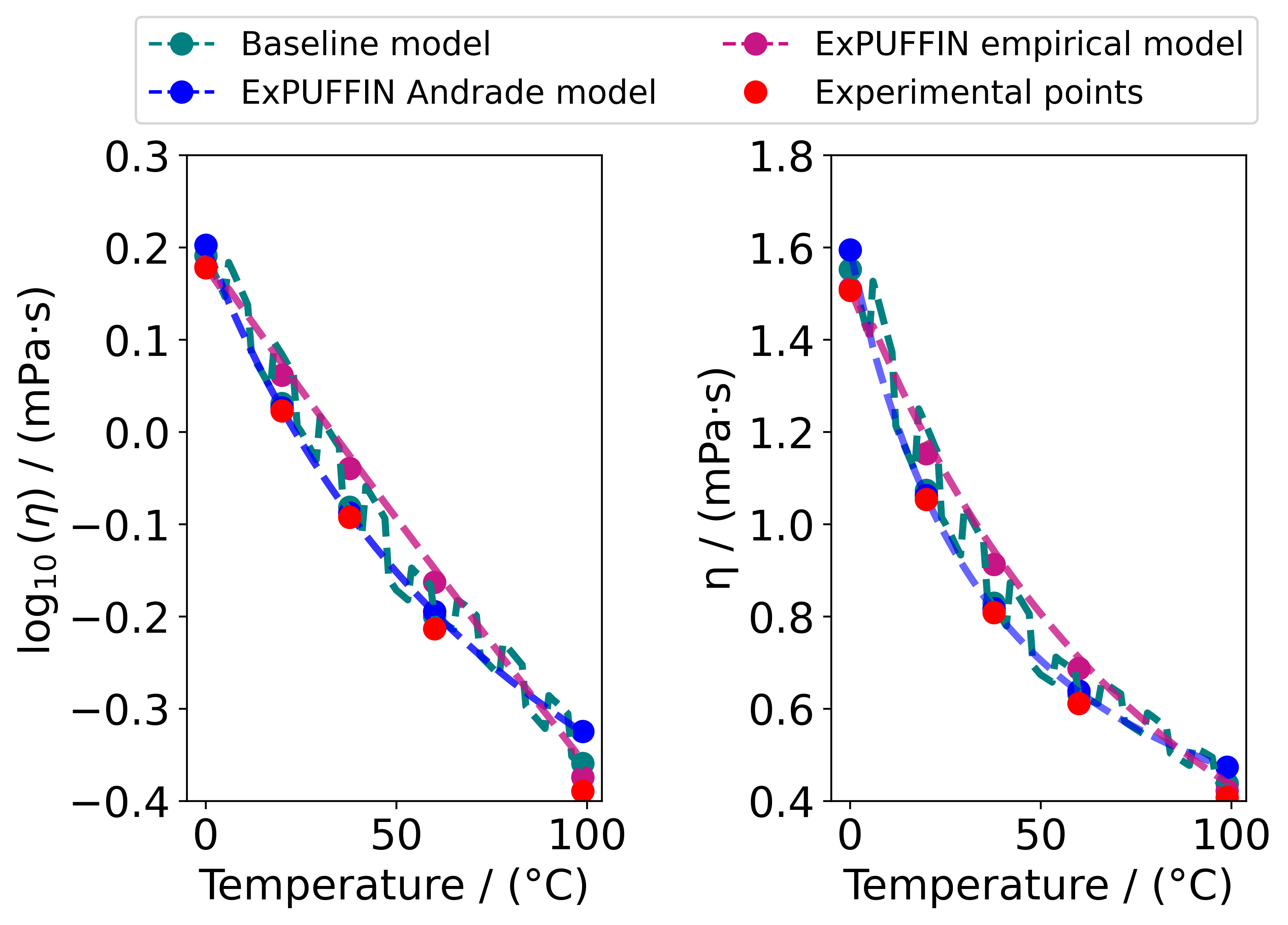}
        \caption{C=CCCCCCCCCC}
    \end{subfigure}
    \hfill
    \begin{subfigure}[b]{0.47\textwidth}
        \centering
        \includegraphics[width=1.0\linewidth]{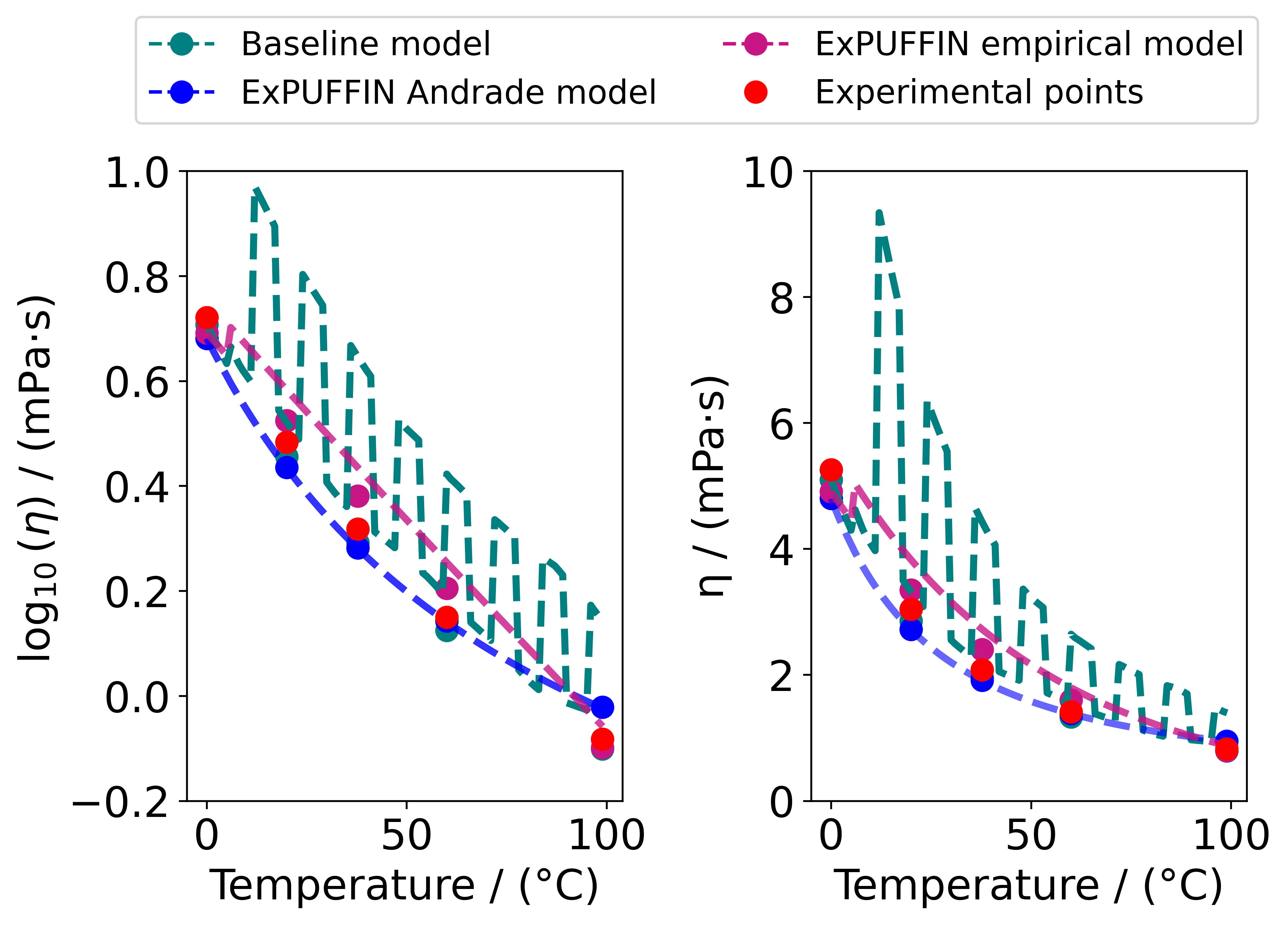}
        \caption{CC(/C(=C/C(C(C)(C)C)(C)C)/C)(C)C}
    \end{subfigure}

    \begin{subfigure}[b]{0.47\textwidth}
        \centering
        \includegraphics[width=1.0\linewidth]{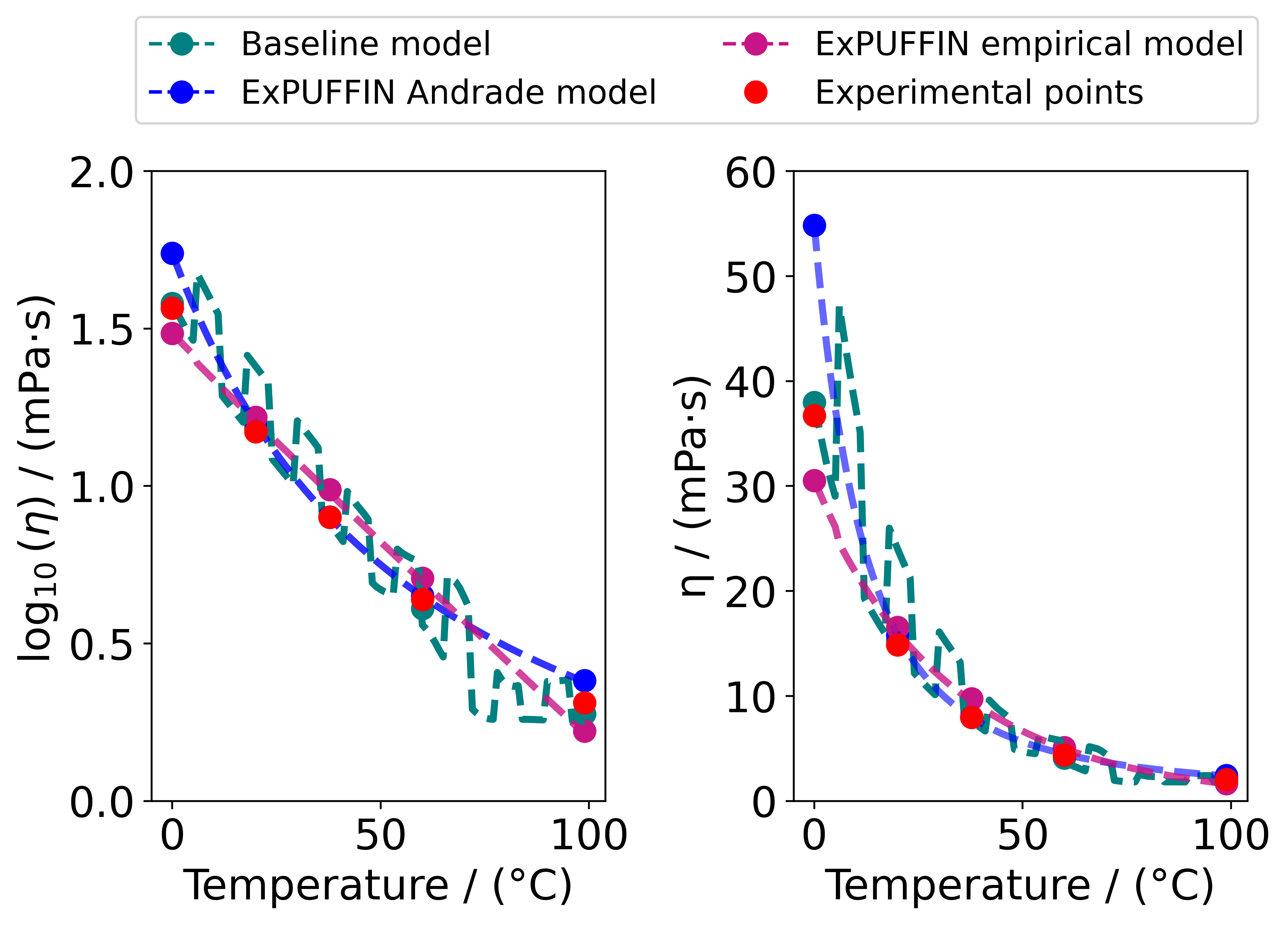}
        \caption{CCCCCCCCCCC(CCCCCCCCCC)CCCCC}
    \end{subfigure}
    \hfill
    \begin{subfigure}[b]{0.47\textwidth}
        \centering
        \includegraphics[width=1.0\linewidth]{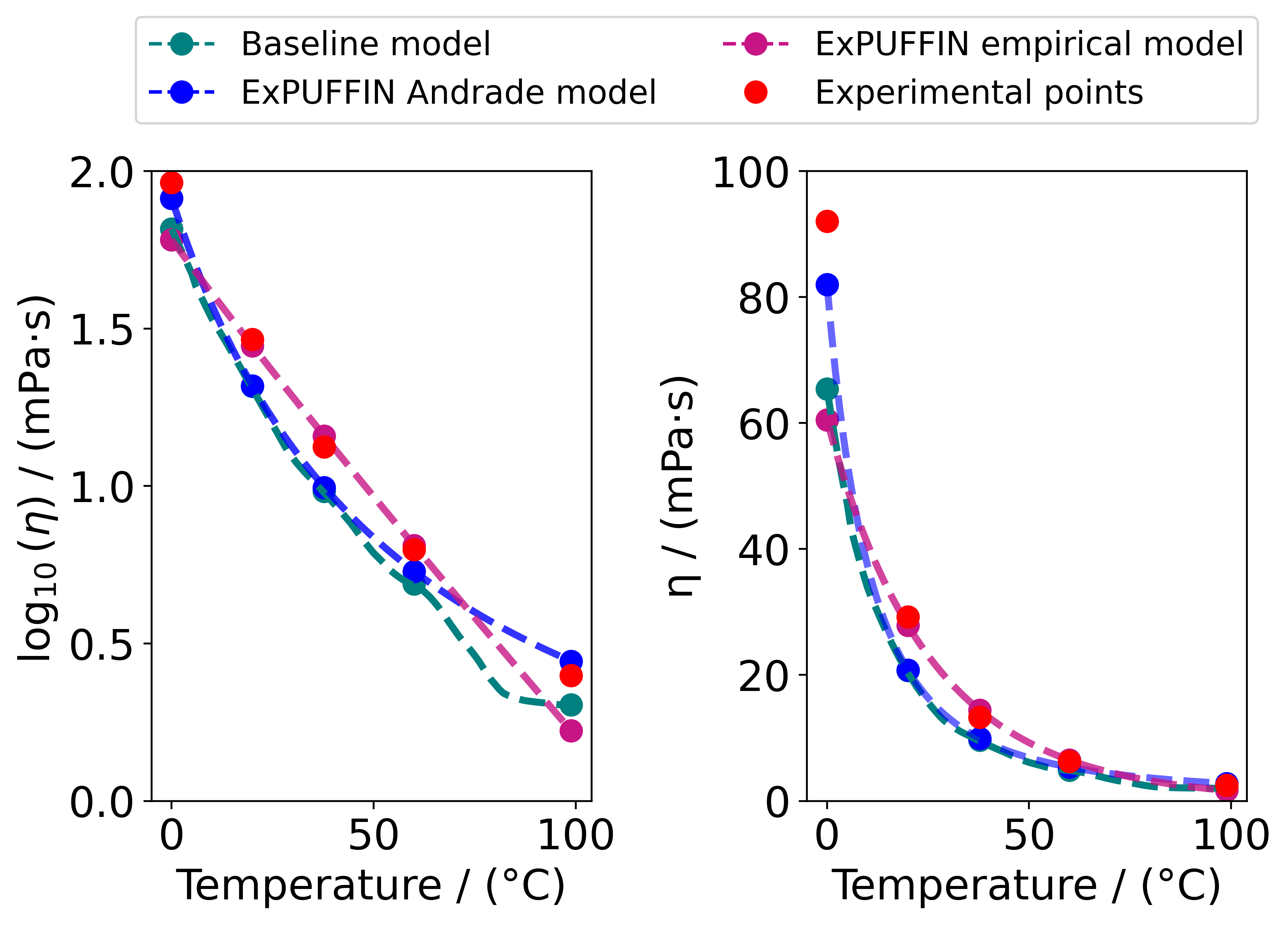}
        \caption{c1c2c(cc(c1CCCC)CCCCCC)CCCC2}
    \end{subfigure}

    \begin{subfigure}[b]{0.47\textwidth}
        \centering
        \includegraphics[width=1.0\linewidth]{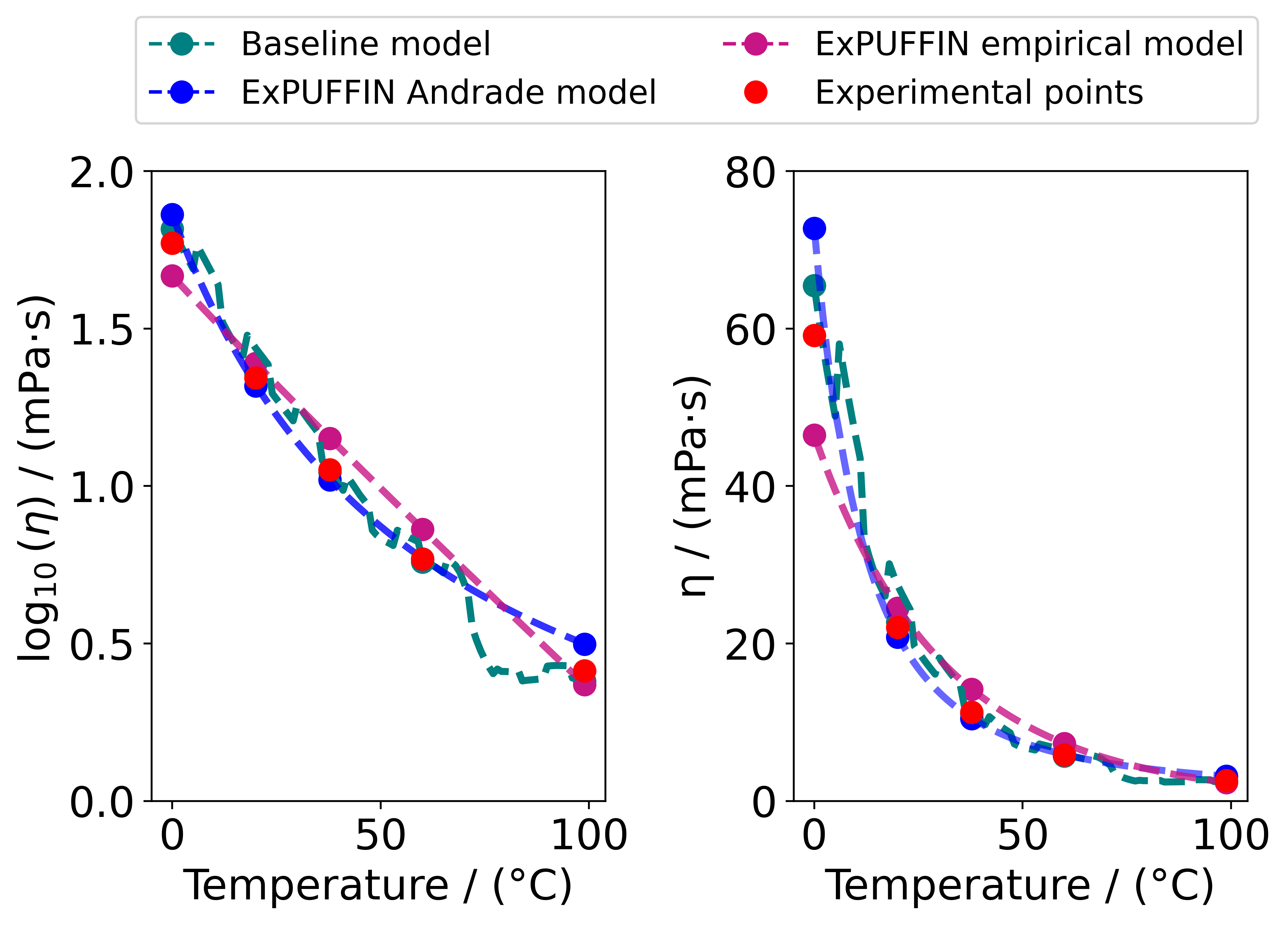}
        \caption{C1(CCCC1)CC(CCCCCCCCCC)CCCCCCCCCC}
    \end{subfigure}
    \hfill
    \begin{subfigure}[b]{0.47\textwidth}
        \centering
        \includegraphics[width=1.0\linewidth]{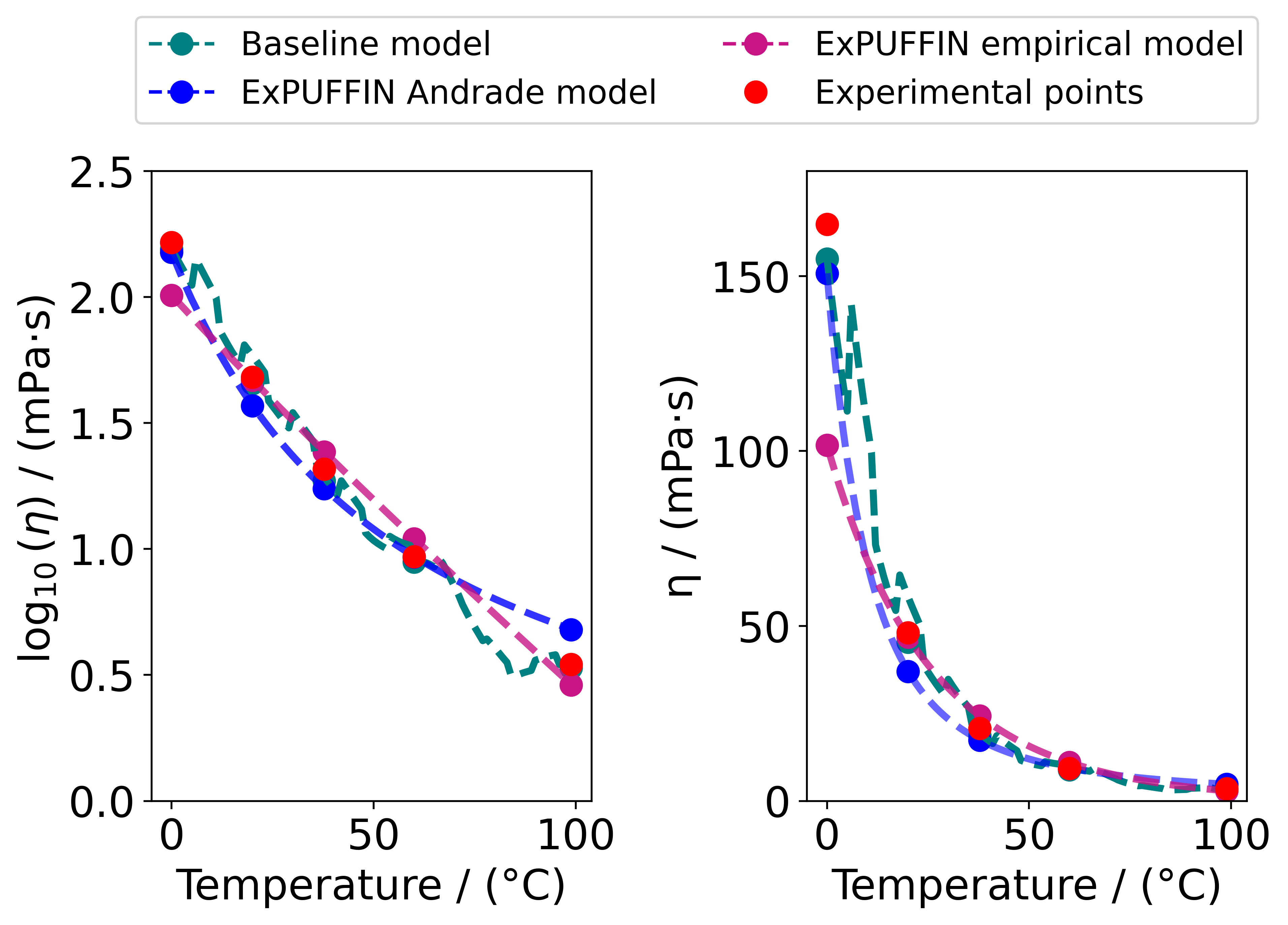}
        \caption{C1C[C@H]2[C@H]3[C@@H](C1)CC[C@@H]1[C@@H]3\\
        {}[C@@H](CCC1)CC2}

    \end{subfigure}

    \caption{Model interpolation of viscosity–temperature behavior for representative hydrocarbons.}
    \label{fig:Interpolation}
\end{figure}

An important part of assessing these models is not just whether they match viscosity at a few temperatures, but whether they can reconstruct the full viscosity–temperature relationship in a smooth and thermodynamically consistent way. To probe this interpolation capability, we evaluated each architecture on a dense temperature grid from $0$ to $100,^{\circ}\mathrm{C}$, within the training range, using a step of $0.5,^{\circ}\mathrm{C}$, and generated continuous viscosity–temperature curves for representative hydrocarbons (Fig.~\ref{fig:Interpolation}). This mirrors practical use in thermophysical property estimation, where viscosity is needed as a continuous function of temperature for simulation, design, and screening, rather than only at the discrete experimental points available in the dataset.

The baseline model matches experimental points reasonably well at the discrete temperatures present in the dataset, but its interpolated trajectories are visibly irregular. For several molecules, the baseline curve oscillates between nearby temperatures. It may locally deviate from a monotonic decrease, despite viscosity being expected to vary smoothly and monotonically with temperature for these systems. These artifacts are not apparent in parity plots or aggregate error metrics, because they occur primarily between training grid points. They are a direct consequence of an unconstrained output head: the network is free to fit the training samples, but it is not structurally required to preserve a physically consistent temperature dependence in between them.

In contrast, both ExPUFFIN variants yield smooth, monotonic viscosity–temperature curves that closely track the experimental trends throughout the full interval. This behavior follows from the proposed inductive-bias architecture: the molecular embedding produced by the GNN controls the viscosity level via a domain-established functional dependence on temperature at the output layer. Therefore, each forward pass inherently respects the expected thermophysical structure, and each backward pass propagates gradients through that same structure, discouraging oscillatory solutions that are mathematically possible but physically implausible. The result is an interpolation that remains coherent even where no measurements exist, reflecting a systematic infusion of domain information at the output stage.

Practically, this means that ExPUFFIN can recover reliable continuous viscosity profiles from comparatively sparse temperature sampling, whereas the baseline requires a denser grid to avoid unphysical wiggles. The interpolation analysis therefore highlights a benefit that is not captured by test-set errors alone: ExPUFFIN preserves baseline-level accuracy at measured points while substantially improving the physical fidelity of the learned temperature dependence between them. This directly supports the central aim of the work, enhancing the reliability of data-driven viscosity prediction through domain-informed inputs and outputs.

\subsection{Extrapolation analysis}

\begin{figure}[h!]
    \centering

    \begin{subfigure}[b]{0.46\textwidth}
        \centering
        \includegraphics[width=\linewidth]{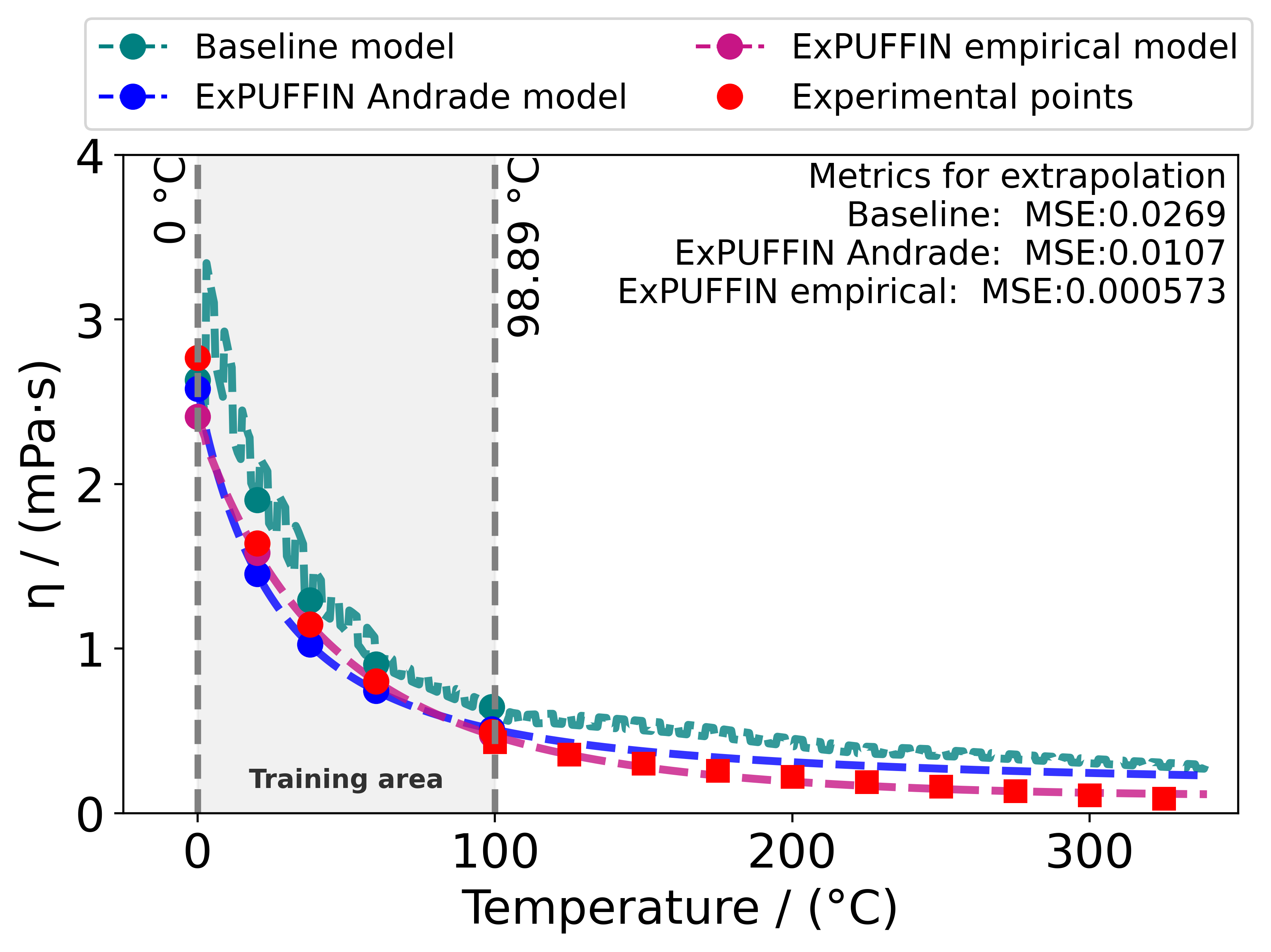}
        \caption{CCCCC(CCCC)CCCC}
    \end{subfigure}
    \hfill
    \begin{subfigure}[b]{0.46\textwidth}
        \centering
        \includegraphics[width=\linewidth]{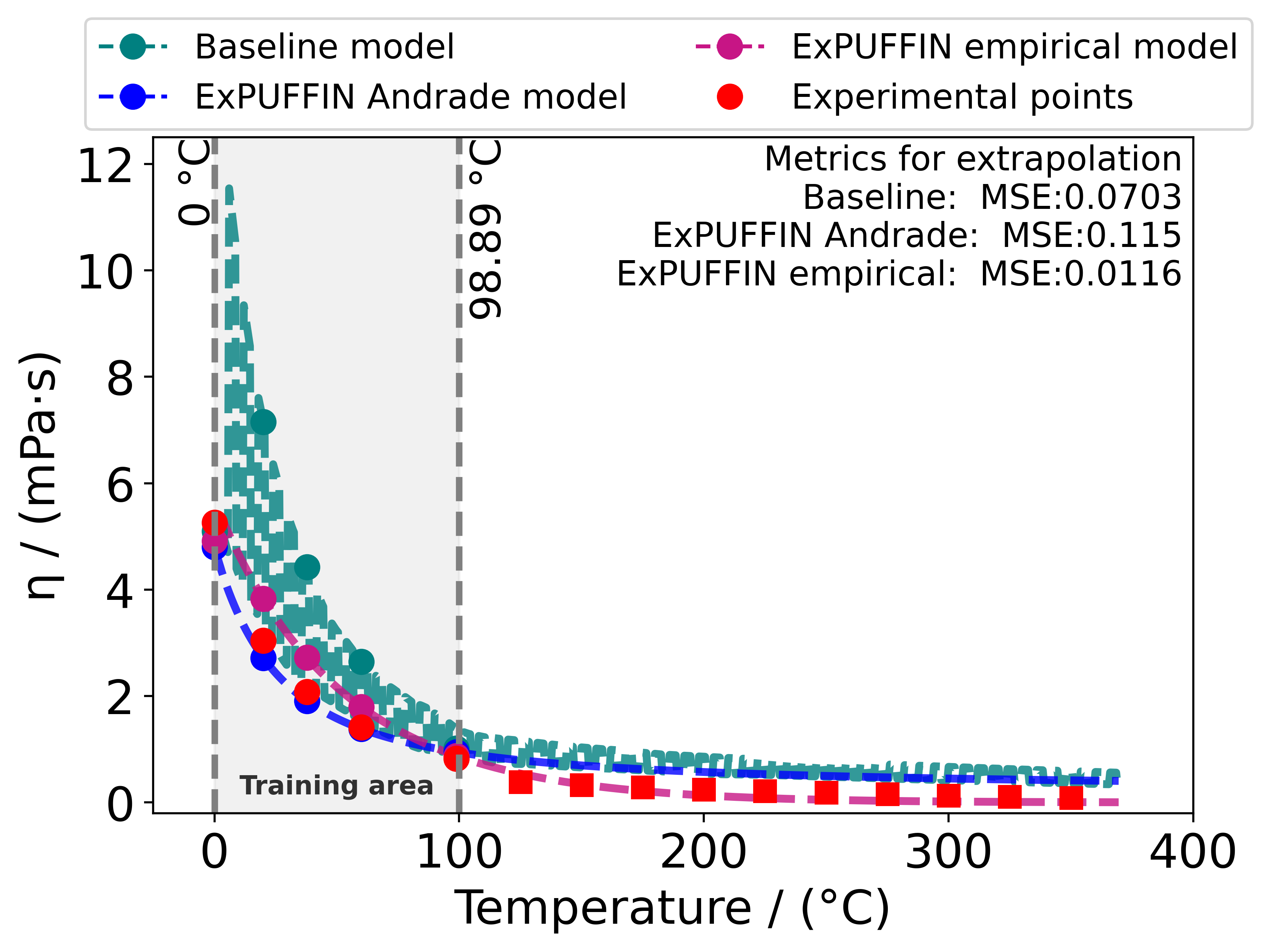}
        \caption{CC(/C(=C/C(C(C)(C)C)(C)C)/C)(C)C}
    \end{subfigure}

    \vspace{0.5em}

    \begin{subfigure}[b]{0.46\textwidth}
        \centering
        \includegraphics[width=\linewidth]{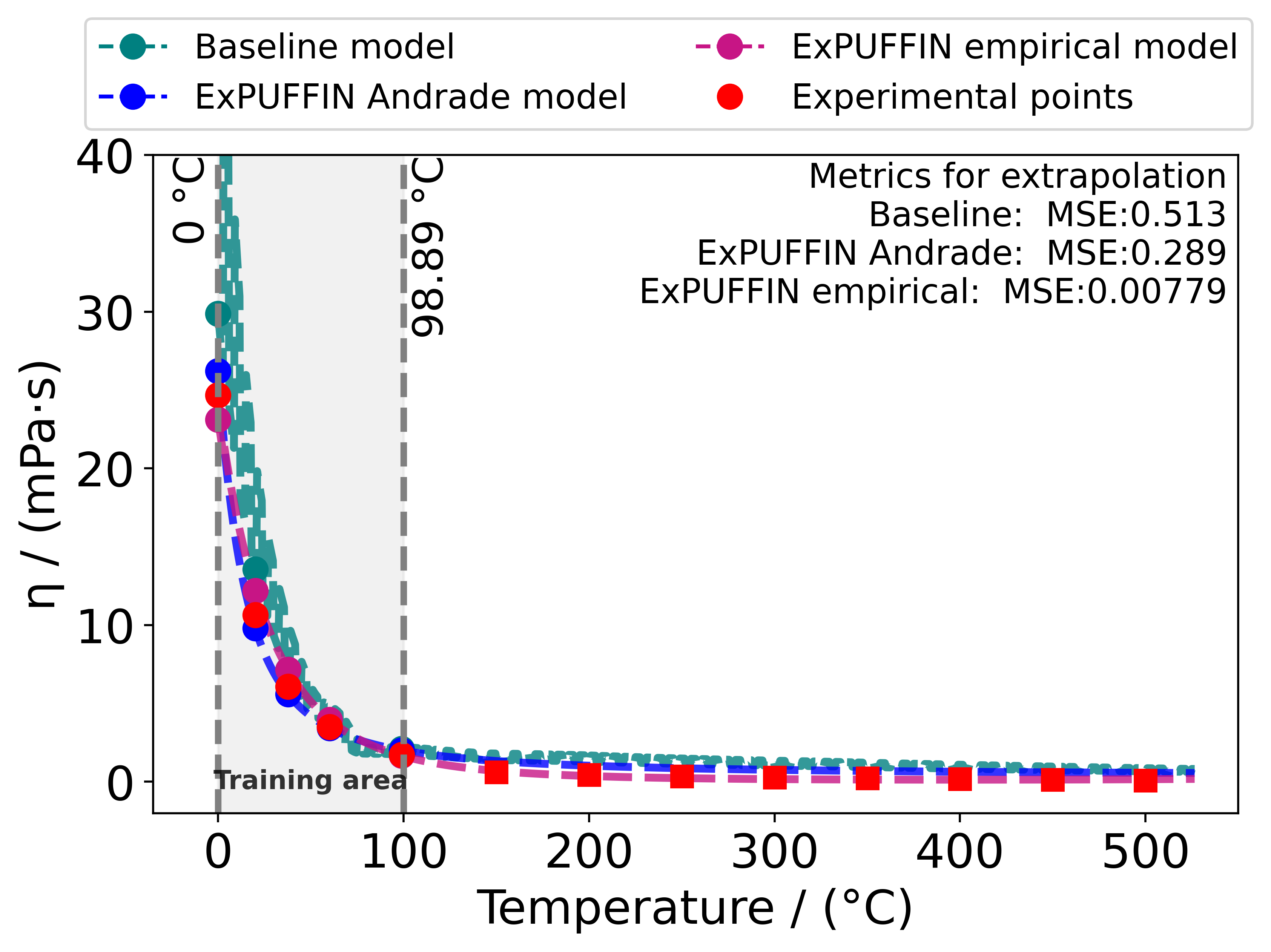}
        \caption{CCCCCCCC=C(CCCCCCCC)CCCCCCCC}
    \end{subfigure}
    \hfill
    \begin{subfigure}[b]{0.46\textwidth}
        \centering
        \includegraphics[width=\linewidth]{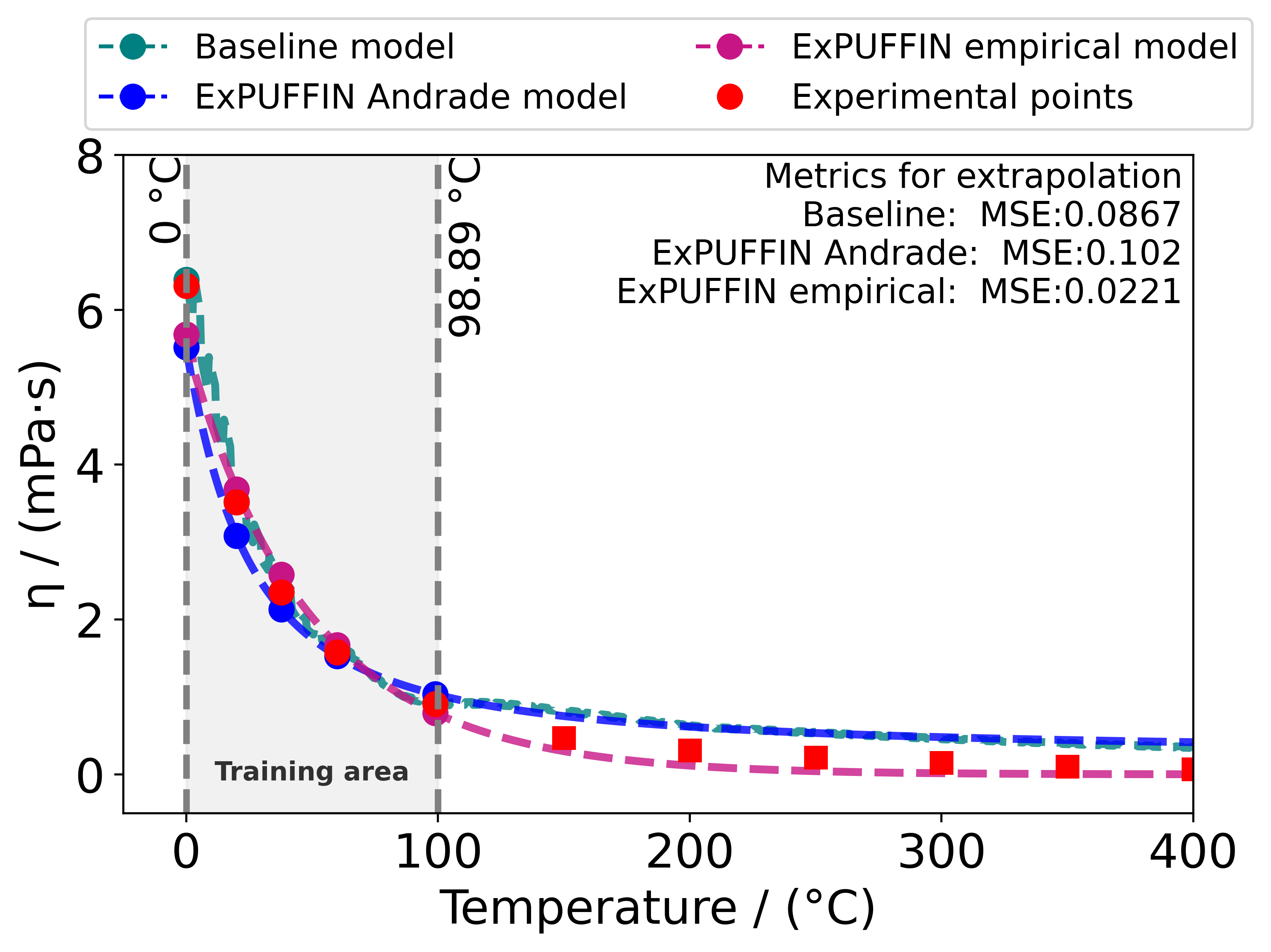}
        \caption{C1CCC(CC1)CCCCCCCC}
    \end{subfigure}

    \begin{subfigure}[b]{0.46\textwidth}
        \centering
        \includegraphics[width=\linewidth]{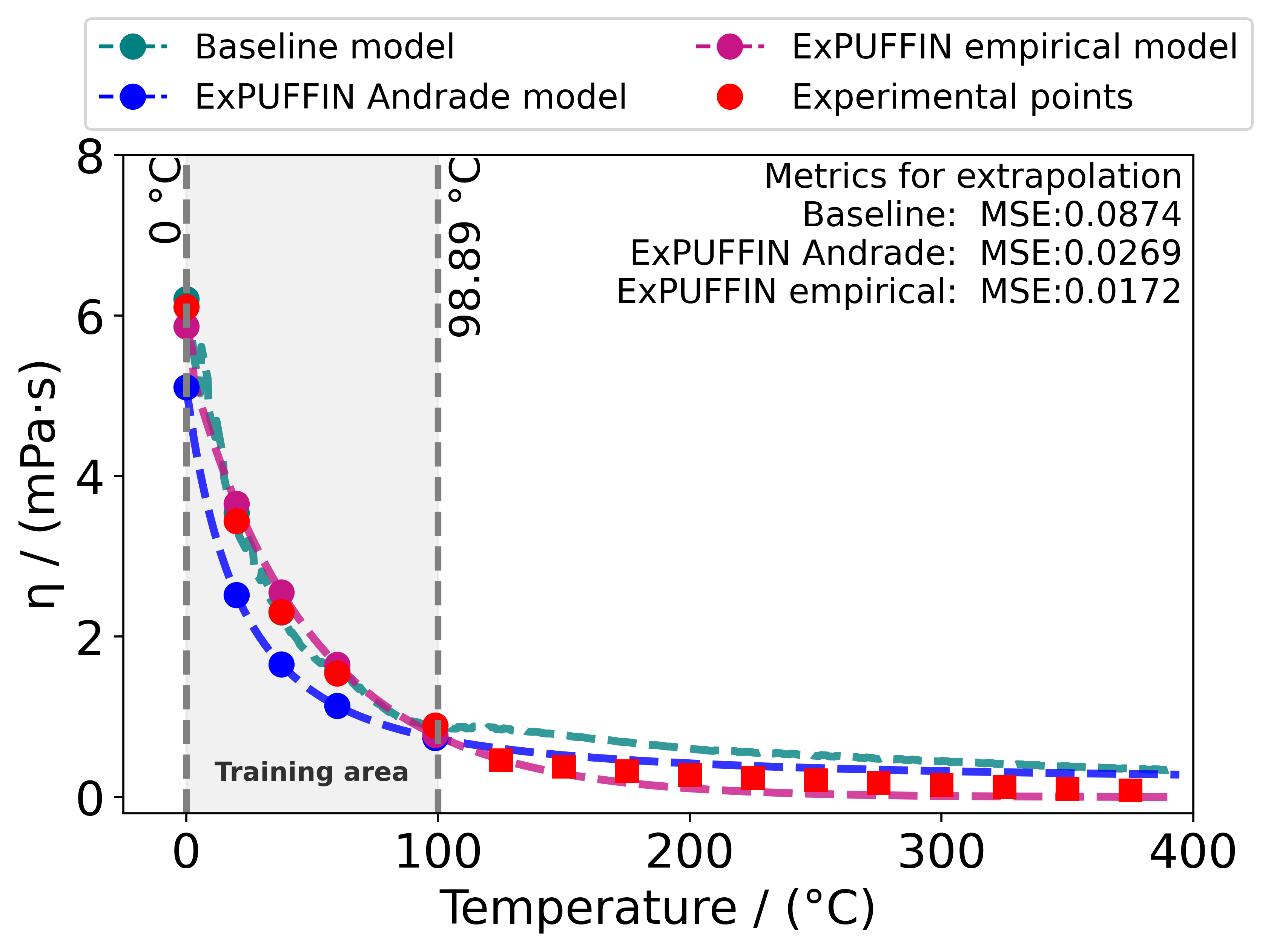}
        \caption{CC(CCCCCCCCCCCCC)C}
    \end{subfigure}
    \hfill
    \begin{subfigure}[b]{0.46\textwidth}
        \centering
        \includegraphics[width=\linewidth]{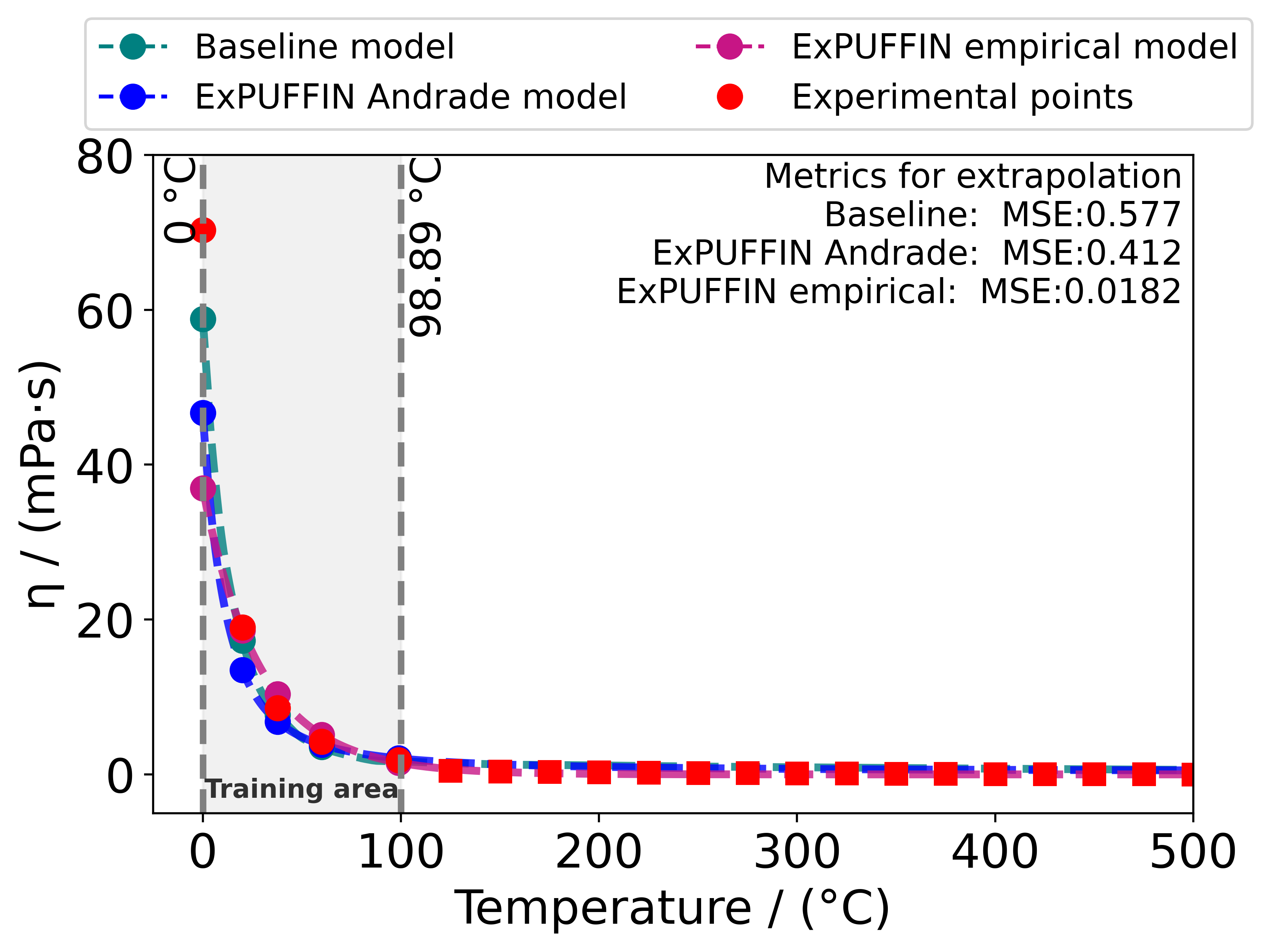}
        \caption{c1ccc(cc1)C(=CCCCCC)c1ccccc1}
    \end{subfigure}

    \caption{Model extrapolation of viscosity–temperature profiles beyond the experimental range.}
    \label{fig:Extrapolation}
\end{figure}

The extrapolation capability of the three architectures was assessed by asking each model to predict viscosity at temperatures well beyond the experimental window used for training. For each representative hydrocarbon in Fig.~\ref{fig:Extrapolation}, viscosity–temperature curves were generated on a dense grid (0.5,$^{\circ}$C resolution) from $T_{\min}$ to $T_{\max}$ (Table~\ref{Tab:extrapolation}). The shaded or dashed boundaries indicate the training region, allowing model behavior to be directly compared inside versus outside the calibration domain. The extrapolated reference points in Fig.~\ref{fig:Extrapolation} were obtained from experimentally identified viscosity correlations reported in the literature (Table~\ref{Tab:extrapolation}; \citep{Yaws2014}).  This analysis is critical because parity plots and test errors only confirm in-range accuracy. They do not reveal whether the learned temperature dependence remains thermodynamically credible when the model is used as practitioners typically use it—namely, for temperatures outside the measured set.

Across all cases, the baseline model quickly loses physical coherence as prediction moves beyond the training range. While it tracks the experimental points within the data window, its extrapolated profiles become oscillatory and may flatten or deviate from the expected monotonic decay of viscosity with temperature. These behaviors are not random noise. They are a structural limitation of the purely data-driven output head. Since the baseline has no enforced functional dependence on temperature, it can fit discrete samples accurately yet still learn a temperature response that is mathematically flexible but thermophysically unreliable once it leaves the region supported by data.

In contrast, both ExPUFFIN variants preserve smooth, monotonic, and thermodynamically plausible viscosity–temperature trends throughout the extended interval. The curves remain stable even hundreds of degrees beyond the training domain, and the extrapolation MSE values shown in Fig.~\ref{fig:Extrapolation} are consistently lower than those of the baseline. This is a direct consequence of the proposed inductive-bias architecture: viscosity is computed using a domain-validated temperature–viscosity functional form in the final layer. During the forward pass, this forces every prediction to obey the known qualitative structure of viscosity with temperature; during backpropagation, the error gradients flow through the same functional constraint, discouraging solutions that fit data locally but violate physics globally. The inductive layer, therefore, acts as a systematic regularizer on the temperature response, not by reducing model flexibility with respect to molecular effects, but by anchoring how these effects manifest across temperature.

The extrapolation results, therefore, close the central argument of this work. Although all three models appear similarly accurate on standard in-range tests, only ExPUFFIN maintains reliable behavior under temperature distribution shift. Embedding domain information in both the molecular input representation and the output mapping enables the model to generalize the shape of the viscosity–temperature relationship in a way that a purely data-driven baseline cannot guarantee. This makes ExPUFFIN better suited for screening novel molecules and operating conditions where extrapolation is unavoidable, and supports our claim that hybridizing GNN structure improves the trustworthiness of viscosity prediction.

\begin{table}[h!]
\centering
\caption{SMILES, molecular formulas, empirical constants and temperature limits for the molecules in Figure~\ref{fig:Extrapolation}.}
\renewcommand{\arraystretch}{1.5}
\scriptsize
\begin{tabularx}{\linewidth}{%
  >{\raggedright\arraybackslash}X  
  c     
  c c c c  
  c c      
}
\hline
\textbf{SMILES} & \textbf{Molecular formula} 
& \textbf{A} & \textbf{B} & \textbf{C} & \textbf{D} 
& $T_{\min}$ (°C) & $T_{\max}$ (°C) \\
\hline

CCCCC(CCCC)CCCC & C$_{13}$H$_{28}$ 
& -9.0731 & 1473.3 & 0.018506 & -1.5$\times10^{-5}$ 
& 61.85 & 343.85 \\

CC(/C(=C/C(C(C)(C)C)(C)C)/C)(C)C & C$_{14}$H$_{28}$
& -7.9996 & 1323.0 & 0.015957 & -1.3$\times10^{-5}$ 
& 80.85 & 372.85 \\

CCCCCCCC=C(CCCCCCCC)CCCCCCCC & C$_{25}$H$_{50}$ 
& -8.6617 & 1676.2 & 0.014924 & -1.0$\times10^{-5}$
& 49.85 & 526.85 \\

C1CCC(CC1)CCCCCCCC & C$_{14}$H$_{28}$ 
& -8.2625 & 1626.0 & 0.014310 & -1.1$\times10^{-5}$
& 26.85 & 413.85 \\

CC(CCCCCCCCCCCCC)C & C$_{16}$H$_{34}$ 
& -10.0624 & 1754.7 & 0.019208 & -1.5$\times10^{-5}$
& 82.85 & 394.85 \\

c1ccc(cc1)C(=CCCCCC)c1ccccc1 & C$_{19}$H$_{22}$
& -8.0811 & 1673.4 & 0.013214 & -9.0$\times10^{-6}$
& 146.85 & 548.85 \\
\hline
\end{tabularx}
\label{Tab:extrapolation}
\end{table}

\section{Conclusions}

This work studied a methodology for predicting the viscosity of pure hydrocarbons, combining graph/based molecular presentations with inductive layers grounded in well-established thermophysical correlations. The results demonstrate that incorporating inductive bias into the ExPUFFIN architectures yields systematic benefits across all evaluation stages, including improved overall accuracy, smoother viscosity/temperature curves, and enhanced stability outside the training region.

The interpolation analysis reinforces these findings: when generating continuous curves across the experimental range, the reference model exhibited structured physical relationships, whereas the model exhibited oscillations and loss of consistency away from the training points. This contrast shows that embedding structured physical relationships into the model output significantly reduces the dependence on densely sampled temperature data, enabling the reconstruction of global trends with less information.

Notably, the extrapolation tests revealed that the ExPUFFIN strategies remain stable at high temperatures far beyond those available in the experimental dataset. For a variety of molecules with distinct structural characteristics, the hybrid predictions preserved the expected thermophysical behavior, while the purely baseline model displayed instabilities and non-physical tendencies. This consistent performance indicates that the proposed approach substantially improves the model's ability to generalize to unobserved conditions.

Overall, the results show that the combination of structural learning and the explicit incorporation of physical knowledge offers a promising path to advance modelling of thermodynamic properties. The ExPUFFIN strategy demonstrates that it is possible to simultaneously enhance accuracy, stability, and thermophysical coherence, thereby paving the way for broader applications in molecular design, process simulation, and the development of compounds for which experimental information is limited or unavailable.

\section*{Data and code availability}
The data and code used in this study are available on the KU Leuven GitLab repository \url{https://gitlab.kuleuven.be/cipt/expuffin_viscosity} or on the Nogueira GitLab NTNU repository
\url{https://github.com/NogueiraGitLabNTNU/ExPUFFIN.git}

\section*{Acknowledgments}
The present work contributes to completion of a sub-project at SUBPRO, a research-based innovation center within Subsea Production and Processing at the Norwegian University of Science and Technology. The authors would like to express their gratitude for the financial support received from SUBPRO, funded by the Research Council of Norway through grant number 237893, major industry partners, and NTNU. UDC, FE, and MEL acknowledge funding from the KU Leuven project HyPro (C3/23/007).



\printcredits

\bibliographystyle{cas-model2-names}

\bibliography{cas-refs}

@article{gani2019,
title = {Group contribution-based property estimation methods: advances and perspectives},
journal = {Current Opinion in Chemical Engineering},
volume = {23},
pages = {184-196},
year = {2019},
note = {Frontiers of Chemical Engineering: Molecular Modeling},
issn = {2211-3398},
doi = {https://doi.org/10.1016/j.coche.2019.04.007},
url = {https://www.sciencedirect.com/science/article/pii/S2211339818300923},
author = {Rafiqul Gani}
}

@article{cardonaetal2021,
author = {Cardona, Luis F. and Forero, Luis A. and Velásquez, Jorge A.},
title = {Extension of a Group Contribution Method to Predict Viscosity Based on Momentum Transport Theory Using a Modified Peng–Robinson EoS},
journal = {Industrial \& Engineering Chemistry Research},
volume = {60},
number = {41},
pages = {14903-14926},
year = {2021},
doi = {10.1021/acs.iecr.1c02146},
URL = {https://doi.org/10.1021/acs.iecr.1c02146},
eprint = {https://doi.org/10.1021/acs.iecr.1c02146} 
}

@article{goussardetal2020,
author = {Goussard, Valentin and Duprat, Fran{\c{c}}ois and Ploix, Jean-Luc and Dreyfus, G{\'e}rard and Nardello-Rataj, V{\'e}ronique and Aubry, Jean-Marie},
title = {A New Machine-Learning Tool for Fast Estimation of Liquid Viscosity. Application to Cosmetic Oils},
journal = {Journal of Chemical Information and Modeling},
volume = {60},
number = {4},
pages = {2012-2023},
year = {2020},
doi = {10.1021/acs.jcim.0c00083},
note ={PMID: 32250628},
URL = { https://doi.org/10.1021/acs.jcim.0c00083},
eprint = {https://doi.org/10.1021/acs.jcim.0c00083}
}

@article{roostaetal2023,
title = {Deep eutectic solvent viscosity prediction by hybrid machine learning and group contribution},
journal = {Journal of Molecular Liquids},
volume = {388},
pages = {122747},
year = {2023},
issn = {0167-7322},
doi = {https://doi.org/10.1016/j.molliq.2023.122747},
url = {https://www.sciencedirect.com/science/article/pii/S0167732223015520},
author = {Ahmadreza Roosta and Reza Haghbakhsh and Ana {Rita C. Duarte} and Sona Raeissi}
}

@incollection{Rittigetalbook2023,
    author = {Rittig, Jan G. and Gao, Qinghe and Dahmen, Manuel and Mitsos, Alexander and Schweidtmann, Artur M.},
    isbn = {978-1-83916-563-4},
    title = {Graph Neural Networks for the Prediction of Molecular Structure–Property Relationships},
    booktitle = {Machine Learning and Hybrid Modelling for Reaction Engineering: Theory and Applications},
    publisher = {Royal Society of Chemistry},
    year = {2023},
    month = {12},
    doi = {10.1039/BK9781837670178-00159},
    url = {https://doi.org/10.1039/BK9781837670178-00159},
}

@article{leenhoutsetal2025,
title = {Property prediction of fuel mixtures using pooled graph neural networks},
journal = {Fuel},
volume = {381},
pages = {133218},
year = {2025},
issn = {0016-2361},
doi = {https://doi.org/10.1016/j.fuel.2024.133218},
url = {https://www.sciencedirect.com/science/article/pii/S0016236124023676},
author = {Roel J. Leenhouts and Tara Larsson and Sebastian Verhelst and Florence H. Vermeire}
}

@article{schweidtmannetal2020,
author = {Schweidtmann, Artur M. and Rittig, Jan G. and K{\"o}nig, Andrea and Grohe, Martin and Mitsos, Alexander and Dahmen, Manuel},
title = {Graph Neural Networks for Prediction of Fuel Ignition Quality},
journal = {Energy \& Fuels},
volume = {34},
number = {9},
pages = {11395-11407},
year = {2020},
doi = {10.1021/acs.energyfuels.0c01533},
URL = {https://doi.org/10.1021/acs.energyfuels.0c01533},
eprint = {https://doi.org/10.1021/acs.energyfuels.0c01533}}

@Article{rittigetal2023,
author ="Rittig, Jan G. and Felton, Kobi C. and Lapkin, Alexei A. and Mitsos, Alexander",
title  ="Gibbs–Duhem-informed neural networks for binary activity coefficient prediction",
journal  ="Digital Discovery",
year  ="2023",
volume  ="2",
issue  ="6",
pages  ="1752-1767",
publisher  ="RSC",
doi  ="10.1039/D3DD00103B",
url  ="http://dx.doi.org/10.1039/D3DD00103B"}

@article{vermeireetal2021,
title = {Transfer learning for solvation free energies: From quantum chemistry to experiments},
journal = {Chemical Engineering Journal},
volume = {418},
pages = {129307},
year = {2021},
issn = {1385-8947},
doi = {https://doi.org/10.1016/j.cej.2021.129307},
url = {https://www.sciencedirect.com/science/article/pii/S1385894721008925},
author = {Florence H. Vermeire and William H. Green}
}

@article{vienasantanaetal2024,
title = {PUFFIN: A path-unifying feed-forward interfaced network for vapor pressure prediction},
journal = {Chemical Engineering Science},
volume = {286},
pages = {119623},
year = {2024},
issn = {0009-2509},
doi = {https://doi.org/10.1016/j.ces.2023.119623},
url = {https://www.sciencedirect.com/science/article/pii/S000925092301179X},
author = {Vinicius Viena Santana and Carine Menezes Rebello and Luana P. Queiroz and Ana Mafalda Ribeiro and Nadia Shardt and Idelfonso B.R. Nogueira}
}

@article{aouichaouietal2023,
title = {Application of interpretable group-embedded graph neural networks for pure compound properties},
journal = {Computers \& Chemical Engineering},
volume = {176},
pages = {108291},
year = {2023},
issn = {0098-1354},
doi = {https://doi.org/10.1016/j.compchemeng.2023.108291},
url = {https://www.sciencedirect.com/science/article/pii/S0098135423001618},
author = {Adem R.N. Aouichaoui and Fan Fan and Jens Abildskov and Gürkan Sin}
}

@article{bilodeauetal2023,
title = {Machine learning for predicting the viscosity of binary liquid mixtures},
journal = {Chemical Engineering Journal},
volume = {464},
pages = {142454},
year = {2023},
issn = {1385-8947},
doi = {https://doi.org/10.1016/j.cej.2023.142454},
url = {https://www.sciencedirect.com/science/article/pii/S1385894723011853},
author = {Camille Bilodeau and Andrei Kazakov and Sukrit Mukhopadhyay and Jillian Emerson and Tom Kalantar and Chris Muzny and Klavs Jensen}
}

@article{schweidtmannetal2024,
title = {A review and perspective on hybrid modeling methodologies},
journal = {Digital Chemical Engineering},
volume = {10},
pages = {100136},
year = {2024},
issn = {2772-5081},
doi = {https://doi.org/10.1016/j.dche.2023.100136},
url = {https://www.sciencedirect.com/science/article/pii/S2772508123000546},
author = {Artur M. Schweidtmann and Dongda Zhang and Moritz {von Stosch}}
}

@article{chewetal2024,
  title={Advancing material property prediction: using physics-informed machine learning models for viscosity},
  author={Chew, Alex K and Sender, Matthew and Kaplan, Zachary and Chandrasekaran, Anand and Chief Elk, Jackson and Browning, Andrea R and Kwak, H Shaun and Halls, Mathew D and Afzal, Mohammad Atif Faiz},
  journal={Journal of Cheminformatics},
  volume={16},
  number={1},
  pages={31},
  year={2024},
  publisher={Springer}
}

@misc{dataset_viscosity,
    title = {panwarp, SupplementaryMaterials},
    year = {2023},
    note = {Accessed: 2023-12-08},
    howpublished = {\url{https://github.com/panwarp/SupplementaryMaterials/tree/main}},
    author = {Pawan Panwar}
}

@article{viscosity_dataset_article,
copyright = {2023 American Chemical Society},
language = {eng},
address = {United States},
author = {Panwar, Pawan and Yang, Quanpeng and Martini, Ashlie},
issn = {1549-9596},
journal = {Journal of chemical information and modeling},
pages = {2760--2774},
volume = {64},
publisher = {American Chemical Society},
number = {7},
year = {2024},
title = {Temperature-Dependent Density and Viscosity Prediction for Hydrocarbons: Machine Learning and Molecular Dynamics Simulations},
}

@article{Weininger1988,
  title = {SMILES,  a chemical language and information system. 1. Introduction to methodology and encoding rules},
  volume = {28},
  ISSN = {1520-5142},
  url = {http://dx.doi.org/10.1021/ci00057a005},
  DOI = {10.1021/ci00057a005},
  number = {1},
  journal = {Journal of Chemical Information and Computer Sciences},
  publisher = {American Chemical Society (ACS)},
  author = {Weininger,  David},
  year = {1988},
  month = feb,
  pages = {31–36}
}

@article{Gutmann_and_Simmons_1952,
    author = {Gutmann, F. and Simmons, L. M.},
    title = {The Temperature Dependence of the Viscosity of Liquids},
    journal = {Journal of Applied Physics},
    volume = {23},
    number = {9},
    pages = {977-978},
    year = {1952},
    month = {09},
    issn = {0021-8979},
    doi = {10.1063/1.1702361},
    url = {https://doi.org/10.1063/1.1702361},
}

@incollection{Cocker_2007,
title = {3 - PHYSICAL PROPERTIES OF LIQUIDS AND GASES},
editor = {A. Kayode Coker},
booktitle = {Ludwig's Applied Process Design for Chemical and Petrochemical Plants (Fourth Edition)},
publisher = {Gulf Professional Publishing},
edition = {Fourth Edition},
address = {Burlington},
pages = {103-132},
year = {2007},
isbn = {978-0-7506-7766-0},
doi = {https://doi.org/10.1016/B978-075067766-0/50010-5},
url = {https://www.sciencedirect.com/science/article/pii/B9780750677660500105},
author = {A. Kayode Coker}
}

@misc{fey2025pyg20scalablelearning,
      title={PyG 2.0: Scalable Learning on Real World Graphs}, 
      author={Matthias Fey and Jinu Sunil and Akihiro Nitta and Rishi Puri and Manan Shah and Blaž Stojanovič and Ramona Bendias and Alexandria Barghi and Vid Kocijan and Zecheng Zhang and Xinwei He and Jan Eric Lenssen and Jure Leskovec},
      year={2025},
      eprint={2507.16991},
      archivePrefix={arXiv},
      primaryClass={cs.LG},
      url={https://arxiv.org/abs/2507.16991}, 
}

@misc{pytorch-ignite,
  author = {V. Fomin and J. Anmol and S. Desroziers and J. Kriss and A. Tejani},
  title = {High-level library to help with training neural networks in PyTorch},
  year = {2020},
  publisher = {GitHub},
  journal = {GitHub repository},
  howpublished = {\url{https://github.com/pytorch/ignite}},
}

@misc{kipf2017gcn,
      title={Semi-Supervised Classification with Graph Convolutional Networks}, 
      author={Thomas N. Kipf and Max Welling},
      year={2017},
      eprint={1609.02907},
      archivePrefix={arXiv},
      primaryClass={cs.LG},
      url={https://arxiv.org/abs/1609.02907}, 
}

@misc{kingma2017adam,
      title={Adam: A Method for Stochastic Optimization}, 
      author={Diederik P. Kingma and Jimmy Ba},
      year={2017},
      eprint={1412.6980},
      archivePrefix={arXiv},
      primaryClass={cs.LG},
      url={https://arxiv.org/abs/1412.6980}, 
}

@misc{cai2021GraphNorm,
      title={GraphNorm: A Principled Approach to Accelerating Graph Neural Network Training}, 
      author={Tianle Cai and Shengjie Luo and Keyulu Xu and Di He and Tie-Yan Liu and Liwei Wang},
      year={2021},
      eprint={2009.03294},
      archivePrefix={arXiv},
      primaryClass={cs.LG},
      url={https://arxiv.org/abs/2009.03294}, 
}

@misc{ioffe2015BatchNorm,
      title={Batch Normalization: Accelerating Deep Network Training by Reducing Internal Covariate Shift}, 
      author={Sergey Ioffe and Christian Szegedy},
      year={2015},
      eprint={1502.03167},
      archivePrefix={arXiv},
      primaryClass={cs.LG},
      url={https://arxiv.org/abs/1502.03167}, 
}

@inbook{Yaws2014, 
  title = {Viscosity of Liquid – Organic Compounds},
  ISBN = {9780323286589},
  url = {http://dx.doi.org/10.1016/B978-0-323-28658-9.00003-2},
  DOI = {10.1016/b978-0-323-28658-9.00003-2},
  booktitle = {Transport Properties of Chemicals and Hydrocarbons},
  publisher = {Elsevier},
  author = {Yaws,  Carl L.},
  year = {2014},
  pages = {131–254}
}

@article{espositohmpi2025, title={Hybrid modelling approaches in process intensification: A thorough review}, volume={217}, rights={All rights reserved}, ISSN={02552701}, DOI={10.1016/j.cep.2025.110496}, journal={Chemical Engineering and Processing - Process Intensification}, author={Esposito, Flora and Di Caprio, Ulderico and Buzzi, Simona and Vermeire, Florence and Leblebici, M. Enis}, year={2025}, month=nov, pages={110496}, language={en} }

@article{santanaetal2023, title={Efficient hybrid modeling and sorption model discovery for non-linear advection-diffusion-sorption systems: A systematic scientific machine learning approach}, volume={282}, ISSN={00092509}, DOI={10.1016/j.ces.2023.119223}, journal={Chemical Engineering Science}, author={Santana, Vinicius V. and Costa, Erbet and Rebello, Carine M. and Ribeiro, Ana Mafalda and Rackauckas, Christopher and Nogueira, Idelfonso B.R.}, year={2023}, month=dec, pages={119223}, language={en} }

@article{nogueiraetal2022, author = {B. R. Nogueira, Idelfonso and V. Santana, Vinicius and Ribeiro, Ana M. and Rodrigues, Alírio E.}, title = {Using scientific machine learning to develop universal differential equation for multicomponent adsorption separation systems}, journal = {The Canadian Journal of Chemical Engineering}, volume = {100}, number = {9}, pages = {2279-2290}, doi = {https://doi.org/10.1002/cjce.24495}, year = {2022}}

@article{dicaprioetal2023, title={Hybrid modelling of a batch separation process}, volume={177}, rights={All rights reserved}, ISSN={00981354}, DOI={10.1016/j.compchemeng.2023.108319}, journal={Computers \& Chemical Engineering}, author={Di Caprio, Ulderico and Wu, Min and Elmaz, Furkan and Wouters, Yentl and Vandervoort, Niels and Anwar, Ali and Mercelis, Siegfried and Waldherr, Steffen and Hellinckx, Peter and Leblebici, M. Enis}, year={2023}, month=sept, pages={108319}, language={en} }

@article{jesperetal_2025, author = {Frandsen, Jesper and Santana, Vinicius V. and Jul-Rasmussen, Peter and Nogueira, Idelfonso B.R. and Huusom, Jakob K. and Gernaey, Krist V. and Abildskov, Jens}, title = {A systematic screening of neural network-based hybrid models of adsorption in chromatography processes}, journal = {AIChE Journal}, volume = {71}, number = {12}, pages = {e70045}, doi = {https://doi.org/10.1002/aic.70045},year = {2025} }

@article{destroetal2020, title = {A hybrid framework for process monitoring: Enhancing data-driven methodologies with state and parameter estimation}, journal = {Journal of Process Control}, volume = {92}, pages = {333-351}, year = {2020}, issn = {0959-1524}, doi = {https://doi.org/10.1016/j.jprocont.2020.06.002}, url = {https://www.sciencedirect.com/science/article/pii/S0959152420302365}, author = {Francesco Destro and Pierantonio Facco and Salvador {García Muñoz} and Fabrizio Bezzo and Massimiliano Barolo}, }

@article{vonstoschhmreview, title={Hybrid semi-parametric modeling in process systems engineering: Past, present and future}, volume={60}, ISSN={00981354}, DOI={10.1016/j.compchemeng.2013.08.008}, journal={Computers and Chemical Engineering}, publisher={Elsevier Ltd}, author={von Stosch, Moritz and Oliveira, Rui and Peres, Joana and Feyo de Azevedo, Sebastião}, year={2014}, pages={86–101} }


\end{document}